\documentclass[twocolumn,english,reprint,aps,prl]{revtex4-2}
\usepackage[T1]{fontenc}
\usepackage[utf8]{inputenc}
\usepackage{geometry}
\usepackage{babel}
\usepackage{verbatim}
\usepackage{refstyle}
\usepackage{amstext}
\usepackage{graphicx}
\usepackage{esint}
\usepackage{amssymb}
\usepackage[dvipsnames]{xcolor}
\usepackage[unicode=true,pdfusetitle,
 bookmarks=true,bookmarksnumbered=false,bookmarksopen=false,
 breaklinks=false,pdfborder={0 0 1},backref=false,colorlinks=false]
 {hyperref}

\makeatletter

\AtBeginDocument{}
\AtBeginDocument{\providecommand\eqref[1]{\ref{eq:#1}}}
\AtBeginDocument{}
\AtBeginDocument{}

\RS@ifundefined{subsecref}
  {\newref{subsec}{name = \RSsectxt}}
  {}
\RS@ifundefined{thmref}
  {\def\RSthmtxt{theorem~}\newref{thm}{name = \RSthmtxt}}
  {}
\RS@ifundefined{lemref}
  {\def\RSlemtxt{lemma~}\newref{lem}{name = \RSlemtxt}}
  {}

\newref{tab}{refcmd = {{Table\,\ref{#1}}}}
\newref{fig}{refcmd = {Figure\,\ref{#1}}}
\newref{eq}{refcmd = {Eq.\,\textup{(\ref{#1})}}}
\newref{sec}{refcmd = {\textsection\ref{#1}}}
\newref{sub}{refcmd = {\textsection\ref{#1}}}
\newref{enu}{refcmd = {\ref{#1}}}
\newref{alg}{refcmd = {Alg.\,\ref{#1}}}

\makeatother

\begin{document}

\title{OmegaNeuron: Applying GravitySpy Similarity Methods to the Search for LIGO Glitch Witnesses}
\author{Bri Aleman$^1$, Derek Davis$^{2,3}$}

\affiliation{$^1$Department of Physics and Astronomy, California State University,
Northridge, Northridge, California 91330, USA \\
$^2$Department of Physics, University of Rhode Island, Kingston, RI 02881, USA \\
$^{3}$LIGO Laboratory, California Institute of Technology, Pasadena, CA 91125, USA}
\begin{abstract}
Gravitational-wave (GW) astronomy has advanced our understanding of compact mergers through instruments like the Laser Interferometer Gravitational-Wave Observatory (LIGO). However, the extreme sensitivity required for these detections makes the instruments susceptible to short-duration transient noise, or glitches, which obscure GW data. Current tools such as Omega Scan and GravitySpy assist in identifying and classifying such noise, but are limited by manual inspection or dependence on large training sets. To address these challenges, we present \textit{OmegaNeuron}, a machine-learning tool that integrates GravitySpy's image similarity methods with Omega Scan's transient analysis to automate the identification of auxiliary channels that witness glitches. Applied to multiple glitch examples, OmegaNeuron consistently highlighted plausible witness channels and showed strong agreement with existing correlation tools, while providing clearer ranking through a quantitative similarity metric. Integrated into the \texttt{gwdetchar} package, OmegaNeuron enables faster analysis that improves glitch witness identification, enhancing both detector sensitivity and the reliability of gravitational-wave observations.
\end{abstract}
\maketitle

\section{Introduction\label{sec:Introduction}}

Since the first gravitational-wave (GW) detection in 2015~\cite{abbott2016GWfromBBH,2016PhRvL.116m1103A}, the Laser Interferometer Gravitational-Wave Observatory (LIGO), together with the Virgo~\cite{acernese2014advanced} and KAGRA~\cite{kagra2019} observatories, has continued to detect merging black holes~\cite{Abbott_2016BBHs, abbott2016GWfromBBH, Abbott_2016_BBHprop, Abbott_2016rateofBBH} and neutron stars~\cite{Abbott_170817, LIGOScientific:2017ync, Abbott_2016rateofBNS} across the universe. However, detections are often limited due to the instrument's sensitivity to transient noise, or \textit{glitches}. This short-lasting noise can originate from a variety of sources like terrestrial vibrations from trains, equipment malfunction, or environmental disturbances such as ravens pecking on the detector's cooling pipes~\cite{Glanzer2023trains, Nuttall:2018xhi, Washimi2021envinjections, Davis:2022dnd, Effler_2015envInfluence, Accadia2010virgoLSnoise}. The noise most significantly obscures transient GWs~\cite{Davis:2022dnd, Nuttall:2018xhi,Abbott_2016charTransient}, which are the only confidently detected GWs to date~\cite{Abbott_2021contGWsfromNS, Riles2023searchContGWs, Piccinni2022statContGWs}. Glitches introduce unwanted noise at a range of frequencies that can mimic or interfere with GW signals~\cite{Powell_2018paramest, Ghonge_2024glitchimpact}, as famously noted in the first detection of a binary neutron star merger~\cite{Abbott_170817,Pankow_2018glitch170817, Torres2020blipnoise}, hindering our ability to extract valuable data to deepen our understanding of gravity and the universe~\cite{wu2024advancing, Macas2022lowlatency, Merritt2021glitchmitig, Udall2025glitchGW191109}. 

Detecting such minuscule fluctuations in space-time requires strain sensitivity on the order of $\sim~10^{-23}/\sqrt{HZ}$ near the most sensitive frequency band of the detectors (100-300 Hz) ~\cite{Nuttall:2018xhi, Martynov_2016sensitivity, Buikema_2020sensO3}. At such sensitivities, the detectors are highly vulnerable to transient noise. The LIGO detector characterization team makes significant effort to identify sources of these glitches, including the use of several thousands auxiliary channels monitored in and around the detector, serving as potential witnesses of the noise~\cite{Nguyen_2021envNoise, Abbott_2020guideToNoise, Walker_2018lassoregress}. We quickly run into issues as the sheer volume of glitches complicates analysis, and human bias is introduced with manual inspection~\cite{Chatterjee2025glitchmatrix, wu2024advancing, Colgan2020envGlitchML}. Within just 50 days of the very first LIGO run (O1), as many as a million glitches were present in the strain data ~\cite{Powell_2018paramest}, and this number is only expected to increase as detector sensitivity increases. Identifying correlations between auxiliary channel and strain channel glitches also proves to be a difficult task for humans to do manually. 

Many approaches have been developed to identify and correlate glitches with their sources, including statistical correlators, machine-learning (ML) frameworks, and noise-injection studies~\cite{Colgan2020envGlitchML,Valdes2022lowlatChar,janssens2025magneticnoiseinject, Glanzer2023O3GSpyDQ,Helmling_Cornell2024autoenvcoupling, Smith2011hveto, huxford2024performanceIDQ, Vazsonyi2023Qtransform, Robinet2020omicron, Powell_2015}. However, these methods have limitations: statistical tools such as \texttt{gwdetchar.omega}~\cite{gwdetchar2025}, which we refer to as Omega Scan, depend on manual inspection of transients. Others, such as the machine-learning based tool \textit{GravitySpy}~\cite{Zevin_2017GSpy, Coughlin:2019ref, soni2021discovering}, require sufficient examples of each glitch to train reliably.

To address these challenges, we introduce OmegaNeuron, a machine-learning based tool that integrates the GravitySpy similarity framework into Omega Scan. OmegaNeuron enables automated identification of auxiliary channels that may have witnessed transient noise coincident with glitches in the strain channel, referred to as \textit{witness channels}. Section II reviews the background of existing pipelines, Section III describes OmegaNeuron's process, Section IV presents case studies including known and unknown glitches, and Section V discusses broader implications and limitations. OmegaNeuron is implemented within the \texttt{gwdetchar} package~\cite{gwdetchar2025}, making it readily available for use by the detector characterization and broader gravitational-wave community.

\section{Background}

When a glitch can be effectively traced to its source, modifications can be made to the interferometer to reduce or the coupling of this noise into the GW strain channel~\cite{Davis2019improvingSens, Pankow:2018qpo, Nuttall:2018xhi, Davis2022subtractingGlitches}. It is generally easier to demonstrate that a signal is a glitch than to confirm it is a GW event, therefore a key focus of detector characterization is to extract evidence that the glitch originates from instrumental or environmental factors~\cite{Nuttall:2018xhi, Davis:2022dnd, Effler_2015envInfluence, Nguyen_2021envNoise, Washimi2021envinjections, Colgan2020envGlitchML}. This process of witness identification and mitigation improves the quality of detected signals, reduces the false-positive rate, and enables more rapid publication of confident GW detections~\cite{LIGO:2021ppb,Smith2011hveto,Hanna2006falsealarm}.

The detector characterization group works on understanding and mitigating transient noise, and several pipelines have been implemented to identify glitch witnesses. Statistical approaches such as HVeto~\cite{Smith2011hveto} and iDQ~\cite{essick2020idq} analyze large datasets to identify couplings between channels, while parameterized injections are used to measure amplitude couplings between environmental sources and the detector~\cite{Washimi2021envinjections}.  Machine-learning methods such as GravitySpy~\cite{Zevin_2017GSpy,Zevin_2024} classify glitches using convolutional neural networks trained on frequency-series data~\cite{Nuttall:2018xhi,Davis:2022dnd, wu2024advancing, George2018DeepTransferLearning}. While effective for common glitches, these supervised learning methods generally require multiple occurrences of a glitch to train the network, making these less suitable for rare events. In contrast, Omega Scan provides detailed summary pages for a single glitch, but it relies heavily on visual inspection of spectrograms and lacks efficiency in automation. Effectiveness can be increased by combining methods to make up where others lack.

\subsection{Omega Scans}

The \texttt{gwdetchar.omega}~\cite{gwdetchar2025} tool, from here on refered to as \textit{Omega Scan}, is a transient identification algorithm that generates visualizations of many data streams on demand. It displays time series data and spectrograms for the strain channel, along with thousands of auxiliary channels that may display coincident noise. Omega Scan also includes a correlation tool based on statistical similarity, but the feature is not widely used in practice because it is often unreliable due to its usage of matched-filtering, which requires prior knowledge of the glitch to work effectively. As a result, Omega Scan is primarily used as a manual inspection tool, providing visual diagnostics but limited automation in determining correlation. 

\subsection{GravitySpy}

GravitySpy~\cite{Zevin_2017GSpy, Coughlin:2019ref, soni2021discovering} utilizes citizen-science to label glitches and train a convolutional neural network (CNN) to classify transient noise from spectrograms. Each transient is represented by four \texttt{gwpy} spectrograms at different durations (0.5\,s, 1.0\,s, 2.0\,s, and 4.0\,s)~\cite{wu2024advancing} to provide short and long timescale context. The labeled glitches serve as training examples for GravitySpy's neural network, which is then applied to classify larger datasets.

For similarity and clustering, GravitySpy uses a deep embedding method called DIRECT~\cite{bahaadini2018DIRECT}. This embedding is trained on pairs of spectrograms to shape a feature space. Within the feature space, examples of the same class cluster together while different classes are pushed further apart. New spectrograms can then be mapped onto this space. This allows us to look for spectrograms that "look alike" without having to retrain the model. The similarity tool currently is used to evaluate similarities between glitch morphologies within the GW strain channel, and categorizes these glitches into clusters~\cite{Coughlin:2019ref}.

In this work, we use the trained CNN as a feature extractor, requiring no re-training of the model. This tool operates in a high-dimensional space, representing each spectrogram as a 200-dimensional feature vector~\cite{Zevin_2024}. We treat the DIRECT output as a channel descriptor and compare strain and auxiliary spectrograms by cosine distance in this feature space. Because the features capture generic data like time-frequency patterns and not channel-specific data, this method transfers cleanly to auxiliary data.

While Omega Scan offers detailed inspection and GravitySpy provides classification and similarity tools, neither fully addresses the challenge of automated correlation for singular or rare glitches. In this context, a "rare" glitch is a transient that has one or few occurrences, lacking the repeated examples required by statistical methods. Such glitches are among the most difficult to diagnose, especially in low-latency analysis. OmegaNeuron integrates the strength of both approaches to enable efficient identification of auxiliary channels that witness rare glitches. 

\begin{figure*}[t!]
    \includegraphics[width=0.75\textwidth]{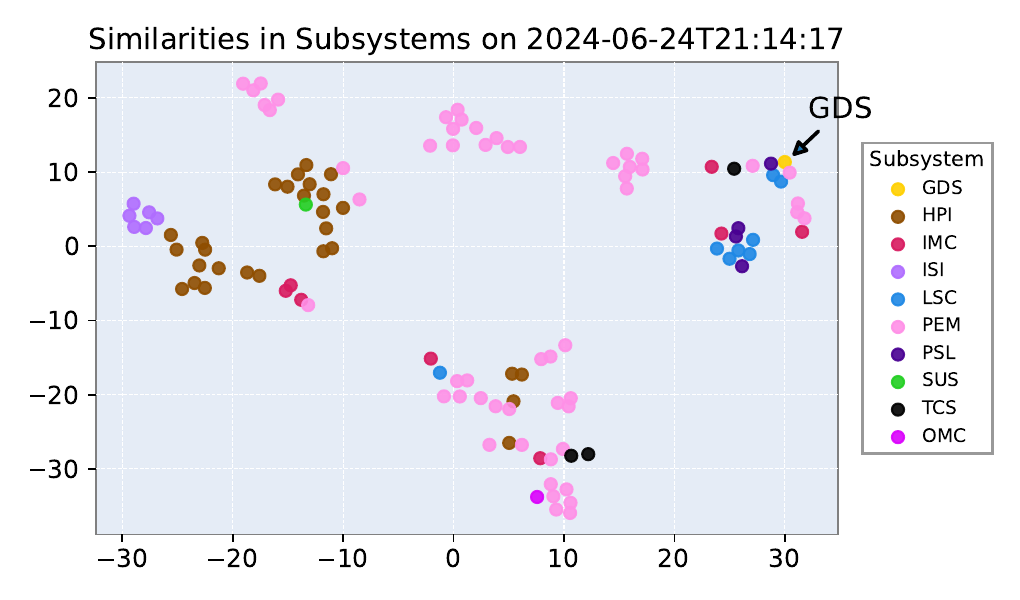}
    
    \caption{t-SNE visualization of the 200-dimensional feature space at GPS time 1403298875.817s. Points are colored by subsystem. Axes are in t-SNE units and not metrically meaningful. Clustering is observed within subsystems. \label{fig:tsne}}
    \end{figure*}

\section{OmegaNeuron}

\subsection{Channel Selection and Setup}

OmegaNeuron is a machine-learning tool designed to identify auxiliary channels correlated with transient noise in LIGO strain data. Previous machine-learning methods often struggle to generalize across noise classes and are limited by rarity of certain glitches~\cite{Coughlin:2019ref, Cuoco2020enhanceGWML}. For single-occurrence glitches, existing tools typically lack the sensitivity or ability to identify witness channels, and will require multiple examples of a glitch to make an assessment. These rare glitches, however, pose some of the more challenging problems in glitch analysis. OmegaNeuron addresses this by analyzing short time windows, enabling effective correlation without relying on large datasets.

The analysis begins with a list of channels obtained by \texttt{gwdetchar} at a single GPS time. These auxiliary channels are selected based on their known potential coupling to the strain channel, including instrumental and environmental monitors~\cite{Smith2011hveto, Jung2022channelCouplings, Essick2013optVetoes, Fiori2020envNoiseBook}. Using \texttt{gwpy}, spectrograms are generated for both the strain and auxiliary channels in windows of 0.5\,s, 1.0\,s, 2.0\,s, and 4.0\,s to mirror the format used in GravitySpy. These multiple windows provide a more complete representation of the glitch, capturing its morphology across different durations. 

Although GravitySpy's similarity tool was initially trained only on GW strain data, its job is to extract generic spectrogram features, which allows it to be applied to auxiliary channels as well. Each spectrogram's features are extracted into a 200-dimensional vector using the trained model, resulting in a high-dimensional feature space. This feature space holds information about each spectrogram's content, such as morphology, frequency, signal-to-noise ratio, etc., which we can then utilize as a similarity database for transient noise.

\subsection{Visualization of Spectrogram Similarity}

To visualize relationships within the high-dimensional feature space, we apply  t-distributed stochastic neighbor embedding (t-SNE)~\cite{vander2008visualizing} to reduce dimensionality. The 2D visualization represents approximate relationships in the full 200-dimensional space, where points close together represent spectrograms with similar features. Because t-SNE preserves local but not global distances, the axes are arbitrary and not metrically meaningful. We therefore use t-SNE only for qualitative inspection, then rely on cosine similarity in the feature space for ranking.

\begin{figure*}[t!]
    \centering
    \vspace{0.5cm}
    \includegraphics[width=0.90\textwidth, height=9cm]{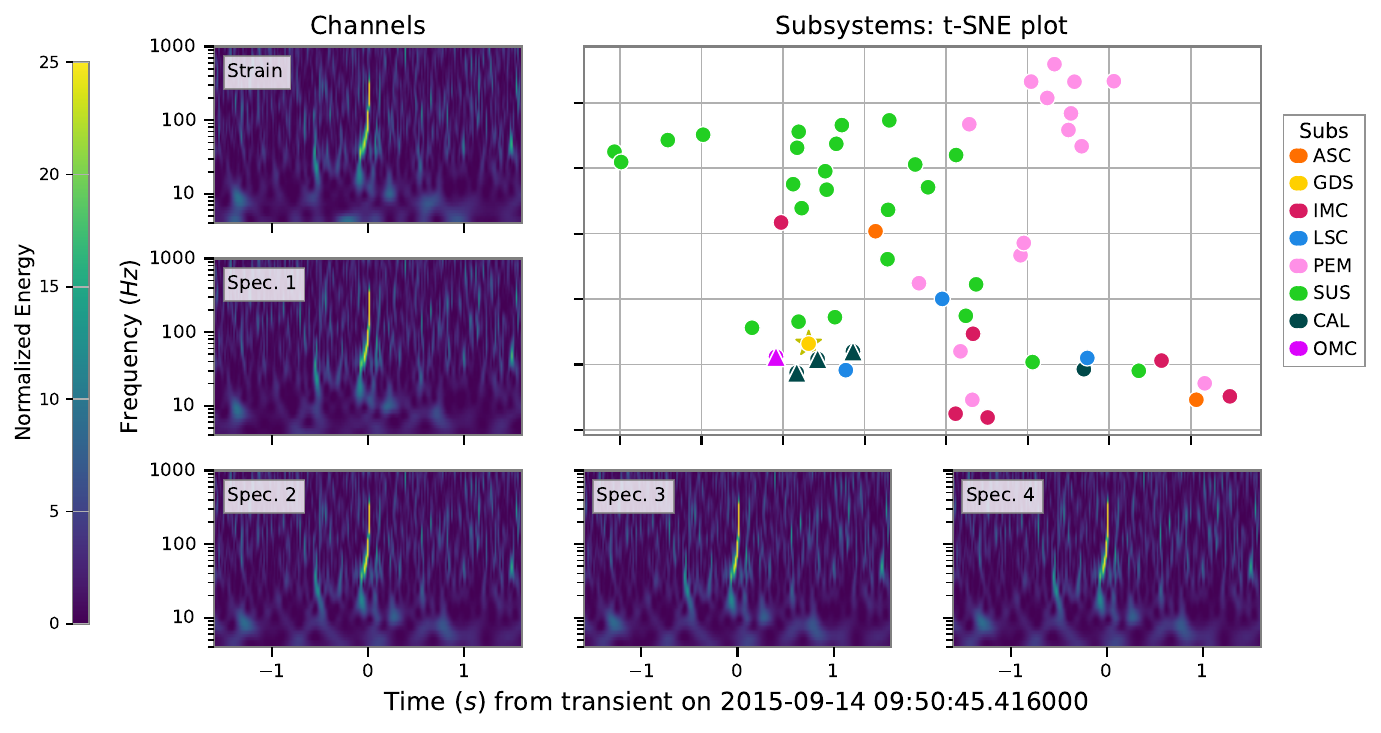}
    
    \caption{OmegaNeuron applied to data from GW150914 (no known glitch). The t-SNE projection shows clustering of auxiliary channels by subsystem; axes are arbitrary. The strain spectrogram is marked with a yellow $\bigstar$, and the top four auxiliary spectrograms plotted in the bottom panels are indicated by $\blacktriangle$. The highest-similarity channels were from the CAL, SUS, and OMC subsystems, all of which are known unsafe couplings. Similarity values above 0.998 demonstrate internal consistency of the metric. A full list of acronyms is provided in Appendix  \ref{app:acronyms}, Table I, and detailed rankings are provided in Appendix \ref{app:ranks} Table II.
    \label{fig:unsafe}}
\end{figure*}

Figure \ref{fig:tsne} shows a t-SNE plot of a rare glitch detected on June 20, 2024. Channels are colored by subsystem (see Appendix \ref{app:acronyms} for a list of acronyms), and clustering is observed within subsystems, consistent with common noise sources or shared coupling. The strain channel is labeled as reference. We are able to confirm feature consistency across related sensors and subsystems, but no channel is prioritized yet in this analysis; all spectrograms are weighed equally, reducing any bias about which channels are significant.

This visualization is used solely for qualitative visualization: it validates GravitySpy's feature extraction through visible subsystem clustering, and highlights any channel groupings of interest. However, t-SNE alone does not provide a quantitative measure of similarity to the strain channel. It is a useful tool for humans to inspect the structure of the high-dimensional data, but it cannot identify witness channels on its own. To do so, we incorporate a cosine distance for further analysis. 

\subsection{Similarity Distance}

To quantify morphological similarity between the strain channel and auxiliary channels, we define a metric called \textit{strain similarity}. Much like in GravitySpy~\cite{Coughlin:2019ref}, this is calculated as the cosine similarity between the 200-dimensional feature vector of the strain channel spectrogram with those of each auxiliary channel spectrograms:

\[\cos(\theta) = \frac{\mathbf{A} \cdot \mathbf{B}}{\Vert \mathbf{A} \Vert \Vert \mathbf{B} \Vert} \] 

where \textbf{A} and \textbf{B} are the feature vectors for the strain and auxiliary channel spectrograms, respectively. Values range from -1, completely dissimilar, to 1, identical. In practice, spectrograms share some generic features, so all observed similarities are positive. Similarity values should be interpreted relative to distribution of all auxiliary channels at a given GPS time. Ranking auxiliary channels by strain similarity identifies those with spectrogram features that are most consistent with the strain glitch morphology, making them candidates for potential glitch witnesses.

 \begin{figure*}[t!]
    \centering
    \vspace{0.5cm}
    \includegraphics[width=0.95\textwidth]{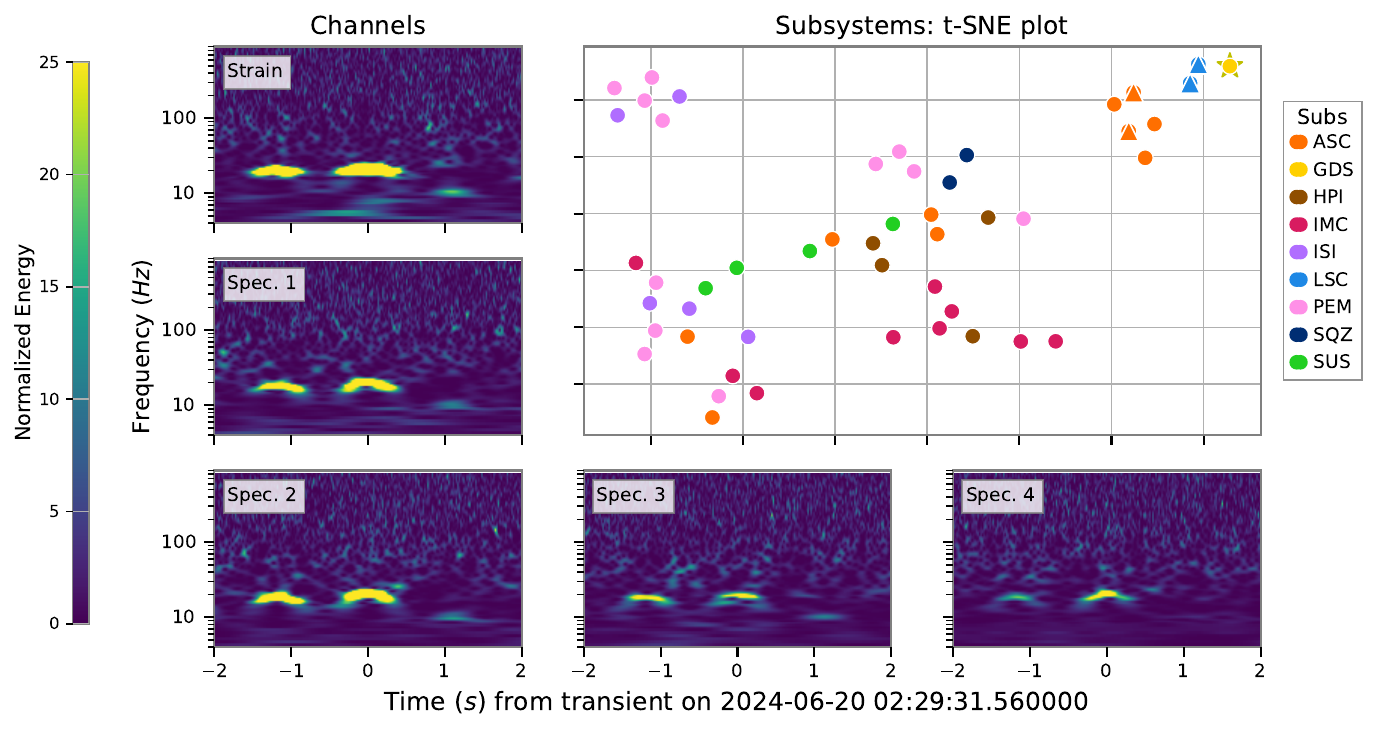}
    \caption{OmegaNeuron applied to a scattered-light glitch. The strain spectrogram is marked with a yellow $\bigstar$, and the four spectrograms shown below are indicated by $\blacktriangle$ in the t-SNE plot. Surrounding clusters in t-SNE plot correspond to the LSC and ASC subsystems, consistent with optical scattering pathways. The two highest-similarity channels closely resemble the strain spectrogram in frequency, duration, and energy, with similarity values above 0.925. Rankings for these and other channels are provided in Appendix \ref{app:ranks} Table III.
    \label{fig:240620}}
\end{figure*}

This analysis relies on two key assumptions. First, we assume the glitch appears simultaneously in both the strain and auxiliary channels. The analysis presented here does not include any delayed coupling beyond that. Second, we assume the glitch looks the same across channels, implying linear coupling. Non-linear or delayed coupling could reduce the effectiveness of the cosine-similarity method. However, these assumptions are generally valid and are supported through correlations observed across many subsystems~\cite{Walker_2018lassoregress, Ajith2014bilinearcoupling,Bose2016BiNonlinearCoupling}. 

To evaluate OmegaNeuron's performance, we apply it to several representative glitch cases. First, we test clean data from signal GW150914 to verify its performance in the absence of glitches. We then examine a well-known scattered-light glitch, which provides a benchmark for comparison against existing correlation tools. Finally, we apply the method to a rare and unclassified glitch to demonstrate its effectiveness for single-occurrence events. These case studies highlight the generalization of OmegaNeuron across different noise classes. 

\section{RESULTS}

\subsection{Case With GW Signal}

We first evaluate OmegaNeuron's behavior in the absence of transient noise by applying it to LIGO Livingston data surrounding GW150914~\cite{2016PhRvL.116m1103A}, the first confirmed GW detection. This event is notable for its clean strain data, with no known glitches disturbing the GW signal. Although OmegaNeuron is designed to identify glitch witness channels, this case provides a verification test of the feature extraction and strain similarity methods. If the tool identifies auxiliary channels with transients resembling the GW signal, this would validate the metrics ability to detect meaningful similarity.

As shown in Figure \ref{fig:unsafe}, OmegaNeuron identified auxiliary spectrograms with high similarity to the GW150914 strain spectrogram. The highest ranked channels belonged to CAL, SUS, and OMC subsystems - each known to be coupled to the strain channel. Several of these are classified as \textit{unsafe} ~\cite{Davis2021detcharO2O3}, meaning they are known to couple to the strain channel and are therefore excluded in the analysis done by glitch mitigation pipelines. However, the appearance of these unsafe channels provides a useful validation: all channels with similarity values above 0.998 showed clear agreement with the strain spectrogram in frequency, duration, and power. A list of acronyms and detailed comparison of OmegaNeuron similarity values and Omega Scan rankings for this glitch is presented in Appendix Tables I and II.

\begin{figure*}[t!]
    \includegraphics[width=0.90\textwidth]{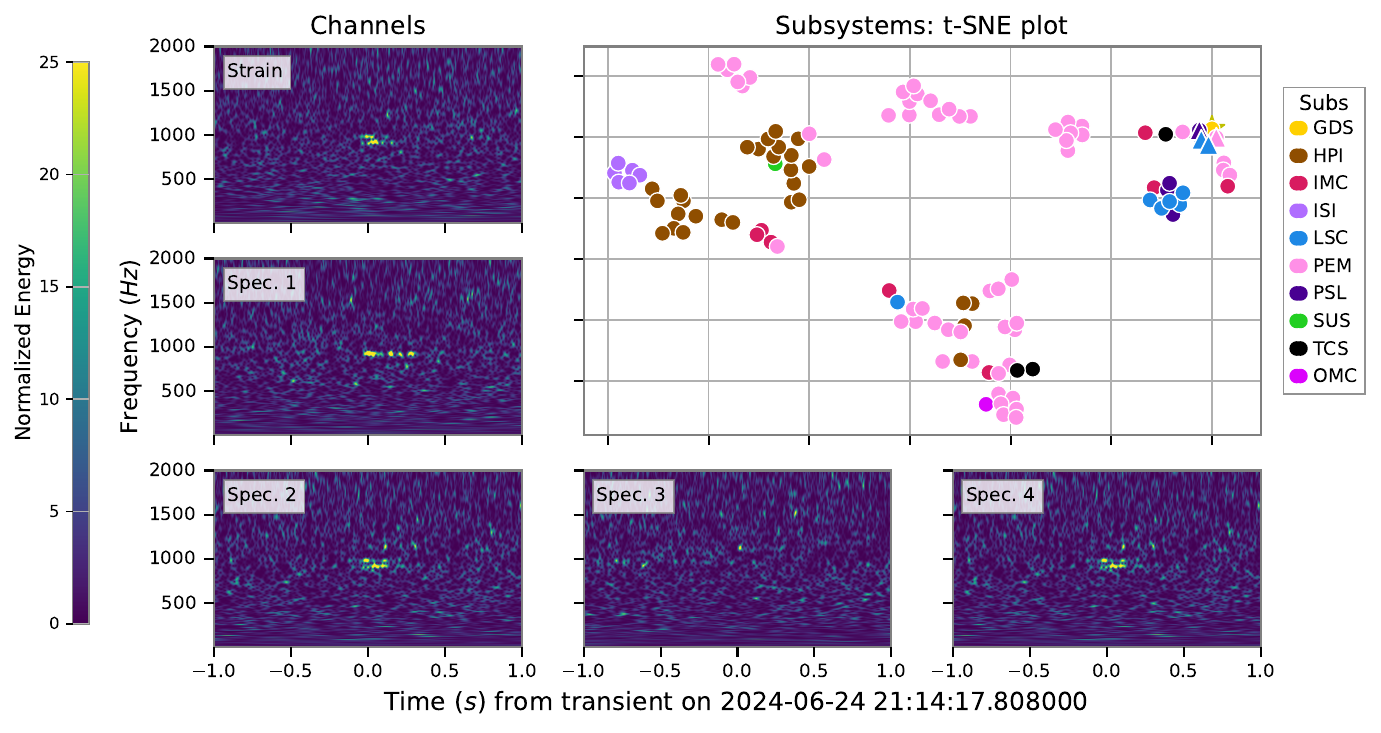}
    \caption{OmegaNeuron applied to an unclassified glitch. The strain spectrogram is marked with a yellow $\bigstar$, and the top four auxiliary spectrograms plotted below are indicated by $\blacktriangle$ in the t-SNE plot. Nearest clusters in the t-SNE projection, consistent with the strain similarity rankings, include PSL, LSC, PEM, and OMC auxiliary channels. The first, second, and fourth highest-similarity spectrograms closely resemble the strain spectrogram in frequency, duration, and energy, with similarity values of 0.9914, 0.9905, and 0.9862, respectively. Rankings for these and other channels are provided in Appendix \ref{app:ranks} Table IV.\label{fig:240624}}
\end{figure*}

Although unsafe channels are not useful for identifying glitch sources, their detection here demonstrates that OmegaNeuron can accurately recognize morphological similarity. All channels with strain similarity values above 0.998 shared visual characteristics with the GW strain spectrogram, confirming the validity of the similarity metric. These values are to be interpreted relative to the similarities of any event, and not general thresholds. Unsafe channels are excluded in following case studies, but this result establishes confidence in our methods. 

\subsection{Case with Known Glitch}

Scattered-light glitches are a persistent and well studied problem in LIGO data~\cite{soni2020scatter, Accadia2010virgoLSnoise,Udall2023bayesianscatter, vinet1996scatteredlight}, especially at the Livingston detector, which we evaluate here. They are among the most common transients, typically originating from light reflecting off surfaces such as beam tubes or other optical components. This scattering introduces low-frequency noise that masks the strain signal. While their morphology is well known, identifying the specific coupling channels has remained challenging. The scattered-light glitch, along with the fast-scattering glitch, are both classes that GravitySpy's neural network is also trained on~\cite{soni2021discovering}.

We applied OmegaNeuron to a representative scattered-light glitch. As shown in Figure \ref{fig:240620}, the t-SNE projection revealed clustering among channels from the LSC and ASC subsystems. These subsystems are directly involved in mirror alignment and sensing, consistent with the expected optical scattering pathways and coupling. Other correlation tools, such as those in Omega Scan, also reported correlations for this glitch, but with different rankings and less accuracy. However, while Omega Scan produced broader statistical correlations across multiple channels, OmegaNeuron assigned the channels the highest similarity values, providing a more interpretable measure of similarity.  A detailed comparison of OmegaNeuron similarity values and Omega Scan rankings for this glitch is presented in Appendix Table III.

\subsection{Case with Unknown Glitch}

Compared to well known glitches such as scattered-light, some glitches do not match any established GravitySpy morphology class~\cite{Coughlin:2019ref, soni2021discovering, Ghonge_2024glitchimpact, Glanzer2023O3GSpyDQ, Davis2022subtractingGlitches, mackenzie2025huntingnewglitches}. We refer to these as "unclassified" glitches. Figure \ref{fig:240624} shows an example of an unclassified glitch. This transient was rare in the data, appearing only a few times within a couple hours, making it difficult for pipelines that rely on repeated occurrences or labeled training examples to determine correlations. In this case, the transient was not assigned to a known glitch category, and did not have enough repeated occurrences to define a new class. Cases such as these provide a useful test of OmegaNeuron's ability to identify witness channels from a single transient. 

By applying OmegaNeuron, one instance of the glitch was enough to accurately identify and point us toward the PSL and LSC subsystems, both of which monitor optical components of the laser. The spectrograms from these channels hold relatively strain similarity, with the top four channels holding a similarity value of >0.980. Both spectrograms visually matched the strain spectrogram in frequency, duration, and energy. Under the assumptions of simultaneity and linear coupling, these channels are strong candidates for having witnessed the glitch. Although this glitch occurred only three times within a single day, OmegaNeuron consistently identified the same witness channels within seconds of analysis. This highlights the tool's effectiveness in rapidly identifying potential sources for rare and unclassified noise transients.

\section{Discussion\label{sec:Discussion}}

OmegaNeuron consistently highlighted plausible auxiliary channels coincident with glitches in the strain channel. Across all tested cases - clean data, scattered-light, and rare, unclassified transients - our method highlighted plausible witness channels, most frequently within optical and environmental subsystems. Although results presented focus on Livingston data, the similarity framework operates only on spectrogram morphology, allowing for straightforward application to other interferometers.

Agreement with independent correlators such as Omega Scan, iDQ, and HVeto further validate our similarity-based ranking. In several cases, OmegaNeuron recovered same or similar coupling channels, but provided clearer prioritization, showing a complement between physical similarity and statistical correlation methods. The present implementation assumes simultaneous and linearly coupled glitches across strain and auxiliary channels, an assumption under which any non-linear or delayed coupling would reduce effectiveness. Extending the method to include time delays or nonlinear coupling will address this limitation.

OmegaNeuron has been integrated into the \texttt{gwdetchar} package, allowing for near real-time analysis within the LIGO detector characterization workflow. The approach can also be adapted for other detectors or re-analysis of archival LIGO data where some noise sources remain ambiguous.

By automating single-glitch correlation and ranking, OmegaNeuron reduces manual inspection time and enhances the reliability of current noise identification methods. These benefits make this tool particularly useful for low latency data-quality vetting, where time is of the essence~\cite{Macas2022lowlatency}. These advances will improve detector sensitivity and data quality, leading to more confident gravitational wave detections.

Applying OmegaNeuron across a range of cases demonstrated that the tool reliably detects meaningful channel correlations. In known glitches, it identified witness channels consistent with previous analyses using other pipelines, and in rare or unclassified cases it provided new insight from single events. These results show OmegaNeuron's ability to provide improved performance over existing correlation-based tools by utilizing this similarity feature space. 

Looking ahead, OmegaNeuron can strengthen other glitch identification pipelines by automating the witness identification process, particularly for rare or new glitch classes. Its speed and generalization make it a great tool for real-time glitch monitoring in current detectors, and for addressing future noise challenges in next-generation detectors~\cite{Garaventa02024nextgen,capote2024DQfornextgen, Srivastava2022CEdetectors}. By improving our ability to identify and mitigate noise at its source, OmegaNeuron advances gravitational-wave detections, improving data quality and confidence in gravitational-wave analysis.

\section*{Acknowledgments}
We thank the Gravity Spy team for discussions and support related to this work. 
A personal acknowledgment is extended to Professors Cristina Cadavid, Chris Theissen, Tyler Luchko, Simon Slutsky, and Abiy Tekola for their support and guidance. 
We also thank Christopher Berry and Jane Glanzer for comments and suggestions during internal review of this work.

This work was supported by the National Science Foundation Research Experience for Undergraduates program through NSF grant PHY-2150027, the Gravity Spy project through NSF grant IIS-2106896, the LIGO Laboratory Summer Undergraduate Research Fellowship program, and the California Institute of Technology Student-Faculty Programs. Continuation of this research was made possible by the Cal-Bridge undergraduate research program.
This work was supported by a grant from the Simons Foundation International [SFI-MPS-SSRFA-00023625, DD].

This material is based upon work supported by NSF’s LIGO Laboratory 
which is a major facility fully funded by the 
National Science Foundation.
LIGO was constructed by the California Institute of Technology 
and Massachusetts Institute of Technology with funding from 
the National Science Foundation, 
and operates under cooperative agreement PHY-1764464. 
Advanced LIGO was built under award PHY-0823459.
The authors are grateful for computational resources provided by the 
LIGO Laboratory and supported by 
National Science Foundation Grants PHY-0757058 and PHY-0823459.

\bibliographystyle{apsrev4-2}
\bibliography{main.bbl}

\begin{thebibliography}{73}%
\makeatletter
\providecommand \@ifxundefined [1]{%
 \@ifx{#1\undefined}
}%
\providecommand \@ifnum [1]{%
 \ifnum #1\expandafter \@firstoftwo
 \else \expandafter \@secondoftwo
 \fi
}%
\providecommand \@ifx [1]{%
 \ifx #1\expandafter \@firstoftwo
 \else \expandafter \@secondoftwo
 \fi
}%
\providecommand \natexlab [1]{#1}%
\providecommand \enquote  [1]{``#1''}%
\providecommand \bibnamefont  [1]{#1}%
\providecommand \bibfnamefont [1]{#1}%
\providecommand \citenamefont [1]{#1}%
\providecommand \href@noop [0]{\@secondoftwo}%
\providecommand \href [0]{\begingroup \@sanitize@url \@href}%
\providecommand \@href[1]{\@@startlink{#1}\@@href}%
\providecommand \@@href[1]{\endgroup#1\@@endlink}%
\providecommand \@sanitize@url [0]{\catcode `\\12\catcode `\$12\catcode `\&12\catcode `\#12\catcode `\^12\catcode `\_12\catcode `\%12\relax}%
\providecommand \@@startlink[1]{}%
\providecommand \@@endlink[0]{}%
\providecommand \url  [0]{\begingroup\@sanitize@url \@url }%
\providecommand \@url [1]{\endgroup\@href {#1}{\urlprefix }}%
\providecommand \urlprefix  [0]{URL }%
\providecommand \Eprint [0]{\href }%
\providecommand \doibase [0]{https://doi.org/}%
\providecommand \selectlanguage [0]{\@gobble}%
\providecommand \bibinfo  [0]{\@secondoftwo}%
\providecommand \bibfield  [0]{\@secondoftwo}%
\providecommand \translation [1]{[#1]}%
\providecommand \BibitemOpen [0]{}%
\providecommand \bibitemStop [0]{}%
\providecommand \bibitemNoStop [0]{.\EOS\space}%
\providecommand \EOS [0]{\spacefactor3000\relax}%
\providecommand \BibitemShut  [1]{\csname bibitem#1\endcsname}%
\let\auto@bib@innerbib\@empty
\bibitem [{\citenamefont {Abbott}\ \emph {et~al.}(2016)\citenamefont {Abbott} \emph {et~al.}}]{abbott2016GWfromBBH}%
  \BibitemOpen
  \bibfield  {author} {\bibinfo {author} {\bibfnamefont {B.~P.}\ \bibnamefont {Abbott}} \emph {et~al.} (\bibinfo {collaboration} {LIGO Scientific Collaboration and Virgo Collaboration}),\ }\href {https://doi.org/10.1103/PhysRevLett.116.061102} {\bibfield  {journal} {\bibinfo  {journal} {Phys. Rev. Lett.}\ }\textbf {\bibinfo {volume} {116}},\ \bibinfo {pages} {061102} (\bibinfo {year} {2016})}\BibitemShut {NoStop}%
\bibitem [{\citenamefont {{Abbott}}\ \emph {et~al.}(2016)\citenamefont {{Abbott}} \emph {et~al.}}]{2016PhRvL.116m1103A}%
  \BibitemOpen
  \bibfield  {author} {\bibinfo {author} {\bibfnamefont {B.~P.}\ \bibnamefont {{Abbott}}} \emph {et~al.},\ }\href {https://doi.org/10.1103/PhysRevLett.116.131103} {\bibfield  {journal} {\bibinfo  {journal} {Physical Review Letters}\ }\textbf {\bibinfo {volume} {116}},\ \bibinfo {eid} {131103} (\bibinfo {year} {2016})},\ \Eprint {https://arxiv.org/abs/1602.03838} {arXiv:1602.03838 [gr-qc]} \BibitemShut {NoStop}%
\bibitem [{\citenamefont {Acernese}\ \emph {et~al.}(2014)\citenamefont {Acernese}, \citenamefont {Agathos}, \citenamefont {Agatsuma}, \citenamefont {Aisa}, \citenamefont {Allemandou}, \citenamefont {Allocca}, \citenamefont {Amarni}, \citenamefont {Astone}, \citenamefont {Balestri}, \citenamefont {Ballardin} \emph {et~al.}}]{acernese2014advanced}%
  \BibitemOpen
  \bibfield  {author} {\bibinfo {author} {\bibfnamefont {F.}~\bibnamefont {Acernese}}, \bibinfo {author} {\bibfnamefont {M.}~\bibnamefont {Agathos}}, \bibinfo {author} {\bibfnamefont {K.}~\bibnamefont {Agatsuma}}, \bibinfo {author} {\bibfnamefont {D.}~\bibnamefont {Aisa}}, \bibinfo {author} {\bibfnamefont {N.}~\bibnamefont {Allemandou}}, \bibinfo {author} {\bibfnamefont {A.}~\bibnamefont {Allocca}}, \bibinfo {author} {\bibfnamefont {J.}~\bibnamefont {Amarni}}, \bibinfo {author} {\bibfnamefont {P.}~\bibnamefont {Astone}}, \bibinfo {author} {\bibfnamefont {G.}~\bibnamefont {Balestri}}, \bibinfo {author} {\bibfnamefont {G.}~\bibnamefont {Ballardin}}, \emph {et~al.},\ }\href@noop {} {\bibfield  {journal} {\bibinfo  {journal} {Classical and Quantum Gravity}\ }\textbf {\bibinfo {volume} {32}},\ \bibinfo {pages} {024001} (\bibinfo {year} {2014})}\BibitemShut {NoStop}%
\bibitem [{\citenamefont {Akutsu}\ \emph {et~al.}(2019)\citenamefont {Akutsu} \emph {et~al.}}]{kagra2019}%
  \BibitemOpen
  \bibfield  {author} {\bibinfo {author} {\bibfnamefont {T.}~\bibnamefont {Akutsu}} \emph {et~al.},\ }\href {https://doi.org/10.1038/s41550-018-0658-y} {\bibfield  {journal} {\bibinfo  {journal} {Nature Astronomy}\ }\textbf {\bibinfo {volume} {3}},\ \bibinfo {pages} {35} (\bibinfo {year} {2019})}\BibitemShut {NoStop}%
\bibitem [{\citenamefont {Abbott}\ \emph {et~al.}(2016{\natexlab{a}})\citenamefont {Abbott} \emph {et~al.}}]{Abbott_2016BBHs}%
  \BibitemOpen
  \bibfield  {author} {\bibinfo {author} {\bibfnamefont {B.~P.}\ \bibnamefont {Abbott}} \emph {et~al.},\ }\bibfield  {journal} {\bibinfo  {journal} {Physical Review X}\ }\textbf {\bibinfo {volume} {6}},\ \href {https://doi.org/10.1103/physrevx.6.041015} {10.1103/physrevx.6.041015} (\bibinfo {year} {2016}{\natexlab{a}})\BibitemShut {NoStop}%
\bibitem [{\citenamefont {Abbott}\ \emph {et~al.}(2016{\natexlab{b}})\citenamefont {Abbott} \emph {et~al.}}]{Abbott_2016_BBHprop}%
  \BibitemOpen
  \bibfield  {author} {\bibinfo {author} {\bibfnamefont {B.~P.}\ \bibnamefont {Abbott}} \emph {et~al.},\ }\bibfield  {journal} {\bibinfo  {journal} {Physical Review Letters}\ }\textbf {\bibinfo {volume} {116}},\ \href {https://doi.org/10.1103/physrevlett.116.241102} {10.1103/physrevlett.116.241102} (\bibinfo {year} {2016}{\natexlab{b}})\BibitemShut {NoStop}%
\bibitem [{\citenamefont {Abbott}\ \emph {et~al.}(2016{\natexlab{c}})\citenamefont {Abbott}, \citenamefont {Abbott}, \citenamefont {Abbott} \emph {et~al.}}]{Abbott_2016rateofBBH}%
  \BibitemOpen
  \bibfield  {author} {\bibinfo {author} {\bibfnamefont {B.~P.}\ \bibnamefont {Abbott}}, \bibinfo {author} {\bibfnamefont {R.}~\bibnamefont {Abbott}}, \bibinfo {author} {\bibfnamefont {T.~D.}\ \bibnamefont {Abbott}}, \emph {et~al.},\ }\href {https://doi.org/10.3847/2041-8205/833/1/l1} {\bibfield  {journal} {\bibinfo  {journal} {The Astrophysical Journal Letters}\ }\textbf {\bibinfo {volume} {833}},\ \bibinfo {pages} {L1} (\bibinfo {year} {2016}{\natexlab{c}})}\BibitemShut {NoStop}%
\bibitem [{\citenamefont {Abbott}\ \emph {et~al.}(2017{\natexlab{a}})\citenamefont {Abbott} \emph {et~al.}}]{Abbott_170817}%
  \BibitemOpen
  \bibfield  {author} {\bibinfo {author} {\bibfnamefont {B.~P.}\ \bibnamefont {Abbott}} \emph {et~al.},\ }\bibfield  {journal} {\bibinfo  {journal} {Physical Review Letters}\ }\textbf {\bibinfo {volume} {119}},\ \href {https://doi.org/10.1103/physrevlett.119.161101} {10.1103/physrevlett.119.161101} (\bibinfo {year} {2017}{\natexlab{a}})\BibitemShut {NoStop}%
\bibitem [{\citenamefont {Abbott}\ \emph {et~al.}(2017{\natexlab{b}})\citenamefont {Abbott} \emph {et~al.}}]{LIGOScientific:2017ync}%
  \BibitemOpen
  \bibfield  {author} {\bibinfo {author} {\bibfnamefont {B.~P.}\ \bibnamefont {Abbott}} \emph {et~al.} (\bibinfo {collaboration} {LIGO Scientific, Virgo, and others}),\ }\href {https://doi.org/10.3847/2041-8213/aa91c9} {\bibfield  {journal} {\bibinfo  {journal} {Astrophys. J. Lett.}\ }\textbf {\bibinfo {volume} {848}},\ \bibinfo {pages} {L12} (\bibinfo {year} {2017}{\natexlab{b}})},\ \Eprint {https://arxiv.org/abs/1710.05833} {arXiv:1710.05833 [astro-ph.HE]} \BibitemShut {NoStop}%
\bibitem [{\citenamefont {Abbott}\ \emph {et~al.}(2016{\natexlab{d}})\citenamefont {Abbott}, \citenamefont {Abbott}, \citenamefont {Abbott} \emph {et~al.}}]{Abbott_2016rateofBNS}%
  \BibitemOpen
  \bibfield  {author} {\bibinfo {author} {\bibfnamefont {B.~P.}\ \bibnamefont {Abbott}}, \bibinfo {author} {\bibfnamefont {R.}~\bibnamefont {Abbott}}, \bibinfo {author} {\bibfnamefont {T.~D.}\ \bibnamefont {Abbott}}, \emph {et~al.},\ }\href {https://doi.org/10.3847/2041-8205/832/2/l21} {\bibfield  {journal} {\bibinfo  {journal} {The Astrophysical Journal Letters}\ }\textbf {\bibinfo {volume} {832}},\ \bibinfo {pages} {L21} (\bibinfo {year} {2016}{\natexlab{d}})}\BibitemShut {NoStop}%
\bibitem [{\citenamefont {Glanzer}\ \emph {et~al.}(2023{\natexlab{a}})\citenamefont {Glanzer}, \citenamefont {Soni}, \citenamefont {Spoon}, \citenamefont {Effler},\ and\ \citenamefont {González}}]{Glanzer2023trains}%
  \BibitemOpen
  \bibfield  {author} {\bibinfo {author} {\bibfnamefont {J.}~\bibnamefont {Glanzer}}, \bibinfo {author} {\bibfnamefont {S.}~\bibnamefont {Soni}}, \bibinfo {author} {\bibfnamefont {J.}~\bibnamefont {Spoon}}, \bibinfo {author} {\bibfnamefont {A.}~\bibnamefont {Effler}},\ and\ \bibinfo {author} {\bibfnamefont {G.}~\bibnamefont {González}},\ }\href {https://doi.org/10.1088/1361-6382/acf01f} {\bibfield  {journal} {\bibinfo  {journal} {Classical and Quantum Gravity}\ }\textbf {\bibinfo {volume} {40}},\ \bibinfo {pages} {195015} (\bibinfo {year} {2023}{\natexlab{a}})}\BibitemShut {NoStop}%
\bibitem [{\citenamefont {Nuttall}(2018)}]{Nuttall:2018xhi}%
  \BibitemOpen
  \bibfield  {author} {\bibinfo {author} {\bibfnamefont {L.~K.}\ \bibnamefont {Nuttall}},\ }\href {https://doi.org/10.1098/rsta.2017.0286} {\bibfield  {journal} {\bibinfo  {journal} {Phil. Trans. Roy. Soc. Lond. A}\ }\textbf {\bibinfo {volume} {376}},\ \bibinfo {pages} {20170286} (\bibinfo {year} {2018})},\ \Eprint {https://arxiv.org/abs/1804.07592} {arXiv:1804.07592 [astro-ph.IM]} \BibitemShut {NoStop}%
\bibitem [{\citenamefont {Washimi}\ \emph {et~al.}(2021)\citenamefont {Washimi}, \citenamefont {Yokozawa}, \citenamefont {Tanaka}, \citenamefont {Itoh}, \citenamefont {Kume},\ and\ \citenamefont {Yokoyama}}]{Washimi2021envinjections}%
  \BibitemOpen
  \bibfield  {author} {\bibinfo {author} {\bibfnamefont {T.}~\bibnamefont {Washimi}}, \bibinfo {author} {\bibfnamefont {T.}~\bibnamefont {Yokozawa}}, \bibinfo {author} {\bibfnamefont {T.}~\bibnamefont {Tanaka}}, \bibinfo {author} {\bibfnamefont {Y.}~\bibnamefont {Itoh}}, \bibinfo {author} {\bibfnamefont {J.}~\bibnamefont {Kume}},\ and\ \bibinfo {author} {\bibfnamefont {J.}~\bibnamefont {Yokoyama}},\ }\href {https://doi.org/10.1088/1361-6382/abf89a} {\bibfield  {journal} {\bibinfo  {journal} {Classical and Quantum Gravity}\ }\textbf {\bibinfo {volume} {38}},\ \bibinfo {pages} {125005} (\bibinfo {year} {2021})}\BibitemShut {NoStop}%
\bibitem [{\citenamefont {Davis}\ and\ \citenamefont {Walker}(2022)}]{Davis:2022dnd}%
  \BibitemOpen
  \bibfield  {author} {\bibinfo {author} {\bibfnamefont {D.}~\bibnamefont {Davis}}\ and\ \bibinfo {author} {\bibfnamefont {M.}~\bibnamefont {Walker}},\ }\href {https://doi.org/10.3390/galaxies10010012} {\bibfield  {journal} {\bibinfo  {journal} {Galaxies}\ }\textbf {\bibinfo {volume} {10}},\ \bibinfo {pages} {12} (\bibinfo {year} {2022})}\BibitemShut {NoStop}%
\bibitem [{\citenamefont {Effler}\ \emph {et~al.}(2015)\citenamefont {Effler}, \citenamefont {Schofield}, \citenamefont {Frolov}, \citenamefont {Gonz\'alez}, \citenamefont {Kawabe}, \citenamefont {Smith}, \citenamefont {Birch},\ and\ \citenamefont {McCarthy}}]{Effler_2015envInfluence}%
  \BibitemOpen
  \bibfield  {author} {\bibinfo {author} {\bibfnamefont {A.}~\bibnamefont {Effler}}, \bibinfo {author} {\bibfnamefont {R.~M.~S.}\ \bibnamefont {Schofield}}, \bibinfo {author} {\bibfnamefont {V.~V.}\ \bibnamefont {Frolov}}, \bibinfo {author} {\bibfnamefont {G.}~\bibnamefont {Gonz\'alez}}, \bibinfo {author} {\bibfnamefont {K.}~\bibnamefont {Kawabe}}, \bibinfo {author} {\bibfnamefont {J.~R.}\ \bibnamefont {Smith}}, \bibinfo {author} {\bibfnamefont {J.}~\bibnamefont {Birch}},\ and\ \bibinfo {author} {\bibfnamefont {R.}~\bibnamefont {McCarthy}},\ }\href {https://doi.org/10.1088/0264-9381/32/3/035017} {\bibfield  {journal} {\bibinfo  {journal} {Classical and Quantum Gravity}\ }\textbf {\bibinfo {volume} {32}},\ \bibinfo {pages} {035017} (\bibinfo {year} {2015})}\BibitemShut {NoStop}%
\bibitem [{\citenamefont {{Accadia}}\ \emph {et~al.}(2010)\citenamefont {{Accadia}}, \citenamefont {{Acernese}}, \citenamefont {{Antonucci}} \emph {et~al.}}]{Accadia2010virgoLSnoise}%
  \BibitemOpen
  \bibfield  {author} {\bibinfo {author} {\bibfnamefont {T.}~\bibnamefont {{Accadia}}}, \bibinfo {author} {\bibfnamefont {F.}~\bibnamefont {{Acernese}}}, \bibinfo {author} {\bibfnamefont {F.}~\bibnamefont {{Antonucci}}}, \emph {et~al.},\ }\href {https://doi.org/10.1088/0264-9381/27/19/194011} {\bibfield  {journal} {\bibinfo  {journal} {Classical and Quantum Gravity}\ }\textbf {\bibinfo {volume} {27}},\ \bibinfo {eid} {194011} (\bibinfo {year} {2010})}\BibitemShut {NoStop}%
\bibitem [{\citenamefont {Abbott}\ \emph {et~al.}(2016{\natexlab{e}})\citenamefont {Abbott}, \citenamefont {Abbott} \emph {et~al.}}]{Abbott_2016charTransient}%
  \BibitemOpen
  \bibfield  {author} {\bibinfo {author} {\bibfnamefont {B.~P.}\ \bibnamefont {Abbott}}, \bibinfo {author} {\bibfnamefont {R.}~\bibnamefont {Abbott}}, \emph {et~al.},\ }\href {https://doi.org/10.1088/0264-9381/33/13/134001} {\bibfield  {journal} {\bibinfo  {journal} {Classical and Quantum Gravity}\ }\textbf {\bibinfo {volume} {33}},\ \bibinfo {pages} {134001} (\bibinfo {year} {2016}{\natexlab{e}})}\BibitemShut {NoStop}%
\bibitem [{\citenamefont {Abbott}\ \emph {et~al.}(2021)\citenamefont {Abbott}, \citenamefont {Abbott}, \citenamefont {Abraham} \emph {et~al.}}]{Abbott_2021contGWsfromNS}%
  \BibitemOpen
  \bibfield  {author} {\bibinfo {author} {\bibfnamefont {R.}~\bibnamefont {Abbott}}, \bibinfo {author} {\bibfnamefont {T.~D.}\ \bibnamefont {Abbott}}, \bibinfo {author} {\bibfnamefont {S.}~\bibnamefont {Abraham}}, \emph {et~al.},\ }\bibfield  {journal} {\bibinfo  {journal} {Physical Review D}\ }\textbf {\bibinfo {volume} {104}},\ \href {https://doi.org/10.1103/physrevd.104.082004} {10.1103/physrevd.104.082004} (\bibinfo {year} {2021})\BibitemShut {NoStop}%
\bibitem [{\citenamefont {Riles}(2023)}]{Riles2023searchContGWs}%
  \BibitemOpen
  \bibfield  {author} {\bibinfo {author} {\bibfnamefont {K.}~\bibnamefont {Riles}},\ }\bibfield  {journal} {\bibinfo  {journal} {Living Reviews in Relativity}\ }\textbf {\bibinfo {volume} {26}},\ \href {https://doi.org/10.1007/s41114-023-00044-3} {10.1007/s41114-023-00044-3} (\bibinfo {year} {2023})\BibitemShut {NoStop}%
\bibitem [{\citenamefont {Piccinni}(2022)}]{Piccinni2022statContGWs}%
  \BibitemOpen
  \bibfield  {author} {\bibinfo {author} {\bibfnamefont {O.~J.}\ \bibnamefont {Piccinni}},\ }\href {https://doi.org/10.3390/galaxies10030072} {\bibfield  {journal} {\bibinfo  {journal} {Galaxies}\ }\textbf {\bibinfo {volume} {10}},\ \bibinfo {pages} {72} (\bibinfo {year} {2022})}\BibitemShut {NoStop}%
\bibitem [{\citenamefont {Powell}(2018)}]{Powell_2018paramest}%
  \BibitemOpen
  \bibfield  {author} {\bibinfo {author} {\bibfnamefont {J.}~\bibnamefont {Powell}},\ }\href {https://doi.org/10.1088/1361-6382/aacf18} {\bibfield  {journal} {\bibinfo  {journal} {Classical and Quantum Gravity}\ }\textbf {\bibinfo {volume} {35}},\ \bibinfo {pages} {155017} (\bibinfo {year} {2018})}\BibitemShut {NoStop}%
\bibitem [{\citenamefont {Ghonge}\ \emph {et~al.}(2024)\citenamefont {Ghonge}, \citenamefont {Brandt}, \citenamefont {Sullivan}, \citenamefont {Millhouse}, \citenamefont {Chatziioannou}, \citenamefont {Clark}, \citenamefont {Littenberg}, \citenamefont {Cornish}, \citenamefont {Hourihane},\ and\ \citenamefont {Cadonati}}]{Ghonge_2024glitchimpact}%
  \BibitemOpen
  \bibfield  {author} {\bibinfo {author} {\bibfnamefont {S.}~\bibnamefont {Ghonge}}, \bibinfo {author} {\bibfnamefont {J.}~\bibnamefont {Brandt}}, \bibinfo {author} {\bibfnamefont {J.}~\bibnamefont {Sullivan}}, \bibinfo {author} {\bibfnamefont {M.}~\bibnamefont {Millhouse}}, \bibinfo {author} {\bibfnamefont {K.}~\bibnamefont {Chatziioannou}}, \bibinfo {author} {\bibfnamefont {J.~A.}\ \bibnamefont {Clark}}, \bibinfo {author} {\bibfnamefont {T.}~\bibnamefont {Littenberg}}, \bibinfo {author} {\bibfnamefont {N.}~\bibnamefont {Cornish}}, \bibinfo {author} {\bibfnamefont {S.}~\bibnamefont {Hourihane}},\ and\ \bibinfo {author} {\bibfnamefont {L.}~\bibnamefont {Cadonati}},\ }\bibfield  {journal} {\bibinfo  {journal} {Physical Review D}\ }\textbf {\bibinfo {volume} {110}},\ \href {https://doi.org/10.1103/physrevd.110.122002} {10.1103/physrevd.110.122002} (\bibinfo {year} {2024})\BibitemShut {NoStop}%
\bibitem [{\citenamefont {Pankow}\ \emph {et~al.}(2018{\natexlab{a}})\citenamefont {Pankow}, \citenamefont {Chatziioannou}, \citenamefont {Chase}, \citenamefont {Littenberg}, \citenamefont {Evans}, \citenamefont {McIver}, \citenamefont {Cornish}, \citenamefont {Haster}, \citenamefont {Kanner}, \citenamefont {Raymond}, \citenamefont {Vitale},\ and\ \citenamefont {Zimmerman}}]{Pankow_2018glitch170817}%
  \BibitemOpen
  \bibfield  {author} {\bibinfo {author} {\bibfnamefont {C.}~\bibnamefont {Pankow}}, \bibinfo {author} {\bibfnamefont {K.}~\bibnamefont {Chatziioannou}}, \bibinfo {author} {\bibfnamefont {E.~A.}\ \bibnamefont {Chase}}, \bibinfo {author} {\bibfnamefont {T.~B.}\ \bibnamefont {Littenberg}}, \bibinfo {author} {\bibfnamefont {M.}~\bibnamefont {Evans}}, \bibinfo {author} {\bibfnamefont {J.}~\bibnamefont {McIver}}, \bibinfo {author} {\bibfnamefont {N.~J.}\ \bibnamefont {Cornish}}, \bibinfo {author} {\bibfnamefont {C.-J.}\ \bibnamefont {Haster}}, \bibinfo {author} {\bibfnamefont {J.}~\bibnamefont {Kanner}}, \bibinfo {author} {\bibfnamefont {V.}~\bibnamefont {Raymond}}, \bibinfo {author} {\bibfnamefont {S.}~\bibnamefont {Vitale}},\ and\ \bibinfo {author} {\bibfnamefont {A.}~\bibnamefont {Zimmerman}},\ }\bibfield  {journal} {\bibinfo  {journal} {Physical Review D}\ }\textbf {\bibinfo {volume} {98}},\ \href {https://doi.org/10.1103/physrevd.98.084016} {10.1103/physrevd.98.084016} (\bibinfo {year}
  {2018}{\natexlab{a}})\BibitemShut {NoStop}%
\bibitem [{\citenamefont {Torres-Forn\'e}\ \emph {et~al.}(2020)\citenamefont {Torres-Forn\'e}, \citenamefont {Cuoco}, \citenamefont {Font},\ and\ \citenamefont {Marquina}}]{Torres2020blipnoise}%
  \BibitemOpen
  \bibfield  {author} {\bibinfo {author} {\bibfnamefont {A.}~\bibnamefont {Torres-Forn\'e}}, \bibinfo {author} {\bibfnamefont {E.}~\bibnamefont {Cuoco}}, \bibinfo {author} {\bibfnamefont {J.~A.}\ \bibnamefont {Font}},\ and\ \bibinfo {author} {\bibfnamefont {A.}~\bibnamefont {Marquina}},\ }\bibfield  {journal} {\bibinfo  {journal} {Physical Review D}\ }\textbf {\bibinfo {volume} {102}},\ \href {https://doi.org/10.1103/physrevd.102.023011} {10.1103/physrevd.102.023011} (\bibinfo {year} {2020})\BibitemShut {NoStop}%
\bibitem [{\citenamefont {Wu}\ \emph {et~al.}(2024)\citenamefont {Wu}, \citenamefont {Zevin}, \citenamefont {Berry}, \citenamefont {Crowston}, \citenamefont {{\O}sterlund}, \citenamefont {Doctor}, \citenamefont {Banagiri}, \citenamefont {Jackson}, \citenamefont {Kalogera},\ and\ \citenamefont {Katsaggelos}}]{wu2024advancing}%
  \BibitemOpen
  \bibfield  {author} {\bibinfo {author} {\bibfnamefont {Y.}~\bibnamefont {Wu}}, \bibinfo {author} {\bibfnamefont {M.}~\bibnamefont {Zevin}}, \bibinfo {author} {\bibfnamefont {C.~P.}\ \bibnamefont {Berry}}, \bibinfo {author} {\bibfnamefont {K.}~\bibnamefont {Crowston}}, \bibinfo {author} {\bibfnamefont {C.}~\bibnamefont {{\O}sterlund}}, \bibinfo {author} {\bibfnamefont {Z.}~\bibnamefont {Doctor}}, \bibinfo {author} {\bibfnamefont {S.}~\bibnamefont {Banagiri}}, \bibinfo {author} {\bibfnamefont {C.~B.}\ \bibnamefont {Jackson}}, \bibinfo {author} {\bibfnamefont {V.}~\bibnamefont {Kalogera}},\ and\ \bibinfo {author} {\bibfnamefont {A.~K.}\ \bibnamefont {Katsaggelos}},\ }\href@noop {} {\bibfield  {journal} {\bibinfo  {journal} {arXiv preprint arXiv:2401.12913}\ } (\bibinfo {year} {2024})}\BibitemShut {NoStop}%
\bibitem [{\citenamefont {Macas}\ \emph {et~al.}(2022)\citenamefont {Macas}, \citenamefont {Pooley}, \citenamefont {Nuttall}, \citenamefont {Davis}, \citenamefont {Dyer}, \citenamefont {Lecoeuche}, \citenamefont {Lyman}, \citenamefont {McIver},\ and\ \citenamefont {Rink}}]{Macas2022lowlatency}%
  \BibitemOpen
  \bibfield  {author} {\bibinfo {author} {\bibfnamefont {R.}~\bibnamefont {Macas}}, \bibinfo {author} {\bibfnamefont {J.}~\bibnamefont {Pooley}}, \bibinfo {author} {\bibfnamefont {L.~K.}\ \bibnamefont {Nuttall}}, \bibinfo {author} {\bibfnamefont {D.}~\bibnamefont {Davis}}, \bibinfo {author} {\bibfnamefont {M.~J.}\ \bibnamefont {Dyer}}, \bibinfo {author} {\bibfnamefont {Y.}~\bibnamefont {Lecoeuche}}, \bibinfo {author} {\bibfnamefont {J.~D.}\ \bibnamefont {Lyman}}, \bibinfo {author} {\bibfnamefont {J.}~\bibnamefont {McIver}},\ and\ \bibinfo {author} {\bibfnamefont {K.}~\bibnamefont {Rink}},\ }\bibfield  {journal} {\bibinfo  {journal} {Physical Review D}\ }\textbf {\bibinfo {volume} {105}},\ \href {https://doi.org/10.1103/physrevd.105.103021} {10.1103/physrevd.105.103021} (\bibinfo {year} {2022})\BibitemShut {NoStop}%
\bibitem [{\citenamefont {Merritt}\ \emph {et~al.}(2021)\citenamefont {Merritt}, \citenamefont {Farr}, \citenamefont {Hur}, \citenamefont {Edelman},\ and\ \citenamefont {Doctor}}]{Merritt2021glitchmitig}%
  \BibitemOpen
  \bibfield  {author} {\bibinfo {author} {\bibfnamefont {J.}~\bibnamefont {Merritt}}, \bibinfo {author} {\bibfnamefont {B.}~\bibnamefont {Farr}}, \bibinfo {author} {\bibfnamefont {R.}~\bibnamefont {Hur}}, \bibinfo {author} {\bibfnamefont {B.}~\bibnamefont {Edelman}},\ and\ \bibinfo {author} {\bibfnamefont {Z.}~\bibnamefont {Doctor}},\ }\bibfield  {journal} {\bibinfo  {journal} {Physical Review D}\ }\textbf {\bibinfo {volume} {104}},\ \href {https://doi.org/10.1103/physrevd.104.102004} {10.1103/physrevd.104.102004} (\bibinfo {year} {2021})\BibitemShut {NoStop}%
\bibitem [{\citenamefont {Udall}\ \emph {et~al.}(2025)\citenamefont {Udall}, \citenamefont {Hourihane}, \citenamefont {Miller}, \citenamefont {Davis}, \citenamefont {Chatziioannou}, \citenamefont {Isi},\ and\ \citenamefont {Deshong}}]{Udall2025glitchGW191109}%
  \BibitemOpen
  \bibfield  {author} {\bibinfo {author} {\bibfnamefont {R.}~\bibnamefont {Udall}}, \bibinfo {author} {\bibfnamefont {S.}~\bibnamefont {Hourihane}}, \bibinfo {author} {\bibfnamefont {S.}~\bibnamefont {Miller}}, \bibinfo {author} {\bibfnamefont {D.}~\bibnamefont {Davis}}, \bibinfo {author} {\bibfnamefont {K.}~\bibnamefont {Chatziioannou}}, \bibinfo {author} {\bibfnamefont {M.}~\bibnamefont {Isi}},\ and\ \bibinfo {author} {\bibfnamefont {H.}~\bibnamefont {Deshong}},\ }\bibfield  {journal} {\bibinfo  {journal} {Physical Review D}\ }\textbf {\bibinfo {volume} {111}},\ \href {https://doi.org/10.1103/physrevd.111.024046} {10.1103/physrevd.111.024046} (\bibinfo {year} {2025})\BibitemShut {NoStop}%
\bibitem [{\citenamefont {Martynov}\ \emph {et~al.}(2016)\citenamefont {Martynov}, \citenamefont {Hall}, \citenamefont {Abbott} \emph {et~al.}}]{Martynov_2016sensitivity}%
  \BibitemOpen
  \bibfield  {author} {\bibinfo {author} {\bibfnamefont {D.}~\bibnamefont {Martynov}}, \bibinfo {author} {\bibfnamefont {E.}~\bibnamefont {Hall}}, \bibinfo {author} {\bibfnamefont {B.}~\bibnamefont {Abbott}}, \emph {et~al.},\ }\bibfield  {journal} {\bibinfo  {journal} {Physical Review D}\ }\textbf {\bibinfo {volume} {93}},\ \href {https://doi.org/10.1103/physrevd.93.112004} {10.1103/physrevd.93.112004} (\bibinfo {year} {2016})\BibitemShut {NoStop}%
\bibitem [{\citenamefont {Buikema}\ \emph {et~al.}(2020)\citenamefont {Buikema}, \citenamefont {Cahillane}, \citenamefont {Mansell} \emph {et~al.}}]{Buikema_2020sensO3}%
  \BibitemOpen
  \bibfield  {author} {\bibinfo {author} {\bibfnamefont {A.}~\bibnamefont {Buikema}}, \bibinfo {author} {\bibfnamefont {C.}~\bibnamefont {Cahillane}}, \bibinfo {author} {\bibfnamefont {G.}~\bibnamefont {Mansell}}, \emph {et~al.},\ }\bibfield  {journal} {\bibinfo  {journal} {Physical Review D}\ }\textbf {\bibinfo {volume} {102}},\ \href {https://doi.org/10.1103/physrevd.102.062003} {10.1103/physrevd.102.062003} (\bibinfo {year} {2020})\BibitemShut {NoStop}%
\bibitem [{\citenamefont {Nguyen}\ \emph {et~al.}(2021)\citenamefont {Nguyen}, \citenamefont {Schofield}, \citenamefont {Effler} \emph {et~al.}}]{Nguyen_2021envNoise}%
  \BibitemOpen
  \bibfield  {author} {\bibinfo {author} {\bibfnamefont {P.}~\bibnamefont {Nguyen}}, \bibinfo {author} {\bibfnamefont {R.~M.~S.}\ \bibnamefont {Schofield}}, \bibinfo {author} {\bibfnamefont {A.}~\bibnamefont {Effler}}, \emph {et~al.},\ }\href {https://doi.org/10.1088/1361-6382/ac011a} {\bibfield  {journal} {\bibinfo  {journal} {Classical and Quantum Gravity}\ }\textbf {\bibinfo {volume} {38}},\ \bibinfo {pages} {145001} (\bibinfo {year} {2021})}\BibitemShut {NoStop}%
\bibitem [{\citenamefont {Abbott}\ \emph {et~al.}(2020)\citenamefont {Abbott}, \citenamefont {Abbott}, \citenamefont {Abbott}, \citenamefont {Abraham} \emph {et~al.}}]{Abbott_2020guideToNoise}%
  \BibitemOpen
  \bibfield  {author} {\bibinfo {author} {\bibfnamefont {B.~P.}\ \bibnamefont {Abbott}}, \bibinfo {author} {\bibfnamefont {R.}~\bibnamefont {Abbott}}, \bibinfo {author} {\bibfnamefont {T.~D.}\ \bibnamefont {Abbott}}, \bibinfo {author} {\bibfnamefont {S.}~\bibnamefont {Abraham}}, \emph {et~al.},\ }\href {https://doi.org/10.1088/1361-6382/ab685e} {\bibfield  {journal} {\bibinfo  {journal} {Classical and Quantum Gravity}\ }\textbf {\bibinfo {volume} {37}},\ \bibinfo {pages} {055002} (\bibinfo {year} {2020})}\BibitemShut {NoStop}%
\bibitem [{\citenamefont {Walker}\ \emph {et~al.}(2018)\citenamefont {Walker}, \citenamefont {Agnew}, \citenamefont {Bidler}, \citenamefont {Lundgren}, \citenamefont {Macedo}, \citenamefont {Macleod}, \citenamefont {Massinger}, \citenamefont {Patane},\ and\ \citenamefont {Smith}}]{Walker_2018lassoregress}%
  \BibitemOpen
  \bibfield  {author} {\bibinfo {author} {\bibfnamefont {M.}~\bibnamefont {Walker}}, \bibinfo {author} {\bibfnamefont {A.~F.}\ \bibnamefont {Agnew}}, \bibinfo {author} {\bibfnamefont {J.}~\bibnamefont {Bidler}}, \bibinfo {author} {\bibfnamefont {A.}~\bibnamefont {Lundgren}}, \bibinfo {author} {\bibfnamefont {A.}~\bibnamefont {Macedo}}, \bibinfo {author} {\bibfnamefont {D.}~\bibnamefont {Macleod}}, \bibinfo {author} {\bibfnamefont {T.~J.}\ \bibnamefont {Massinger}}, \bibinfo {author} {\bibfnamefont {O.}~\bibnamefont {Patane}},\ and\ \bibinfo {author} {\bibfnamefont {J.~R.}\ \bibnamefont {Smith}},\ }\href {https://doi.org/10.1088/1361-6382/aae593} {\bibfield  {journal} {\bibinfo  {journal} {Classical and Quantum Gravity}\ }\textbf {\bibinfo {volume} {35}},\ \bibinfo {pages} {225002} (\bibinfo {year} {2018})}\BibitemShut {NoStop}%
\bibitem [{\citenamefont {Chatterjee}\ and\ \citenamefont {Jani}(2025)}]{Chatterjee2025glitchmatrix}%
  \BibitemOpen
  \bibfield  {author} {\bibinfo {author} {\bibfnamefont {C.}~\bibnamefont {Chatterjee}}\ and\ \bibinfo {author} {\bibfnamefont {K.}~\bibnamefont {Jani}},\ }\href {https://doi.org/10.3847/1538-4357/adbb66} {\bibfield  {journal} {\bibinfo  {journal} {The Astrophysical Journal}\ }\textbf {\bibinfo {volume} {982}},\ \bibinfo {pages} {102} (\bibinfo {year} {2025})}\BibitemShut {NoStop}%
\bibitem [{\citenamefont {Colgan}\ \emph {et~al.}(2020)\citenamefont {Colgan}, \citenamefont {Corley}, \citenamefont {Lau}, \citenamefont {Bartos}, \citenamefont {Wright}, \citenamefont {M\'arka},\ and\ \citenamefont {M\'arka}}]{Colgan2020envGlitchML}%
  \BibitemOpen
  \bibfield  {author} {\bibinfo {author} {\bibfnamefont {R.~E.}\ \bibnamefont {Colgan}}, \bibinfo {author} {\bibfnamefont {K.~R.}\ \bibnamefont {Corley}}, \bibinfo {author} {\bibfnamefont {Y.}~\bibnamefont {Lau}}, \bibinfo {author} {\bibfnamefont {I.}~\bibnamefont {Bartos}}, \bibinfo {author} {\bibfnamefont {J.~N.}\ \bibnamefont {Wright}}, \bibinfo {author} {\bibfnamefont {Z.}~\bibnamefont {M\'arka}},\ and\ \bibinfo {author} {\bibfnamefont {S.}~\bibnamefont {M\'arka}},\ }\bibfield  {journal} {\bibinfo  {journal} {Physical Review D}\ }\textbf {\bibinfo {volume} {101}},\ \href {https://doi.org/10.1103/physrevd.101.102003} {10.1103/physrevd.101.102003} (\bibinfo {year} {2020})\BibitemShut {NoStop}%
\bibitem [{\citenamefont {Valdes}\ \emph {et~al.}(2022)\citenamefont {Valdes}, \citenamefont {Hines}, \citenamefont {Nelson}, \citenamefont {Zhang},\ and\ \citenamefont {Guzman}}]{Valdes2022lowlatChar}%
  \BibitemOpen
  \bibfield  {author} {\bibinfo {author} {\bibfnamefont {G.}~\bibnamefont {Valdes}}, \bibinfo {author} {\bibfnamefont {A.}~\bibnamefont {Hines}}, \bibinfo {author} {\bibfnamefont {A.}~\bibnamefont {Nelson}}, \bibinfo {author} {\bibfnamefont {Y.}~\bibnamefont {Zhang}},\ and\ \bibinfo {author} {\bibfnamefont {F.}~\bibnamefont {Guzman}},\ }\bibfield  {journal} {\bibinfo  {journal} {Applied Physics Letters}\ }\textbf {\bibinfo {volume} {121}},\ \href {https://doi.org/10.1063/5.0122495} {10.1063/5.0122495} (\bibinfo {year} {2022})\BibitemShut {NoStop}%
\bibitem [{\citenamefont {Janssens}\ \emph {et~al.}(2025)\citenamefont {Janssens}, \citenamefont {Lawrence}, \citenamefont {Effler}, \citenamefont {Schofield}, \citenamefont {Lalleman}, \citenamefont {Betzwieser}, \citenamefont {Christensen}, \citenamefont {Coughlin}, \citenamefont {Driggers}, \citenamefont {Helmling-Cornell}, \citenamefont {O'Hanlon}, \citenamefont {Quintero}, \citenamefont {Reis},\ and\ \citenamefont {van Remortel}}]{janssens2025magneticnoiseinject}%
  \BibitemOpen
  \bibfield  {author} {\bibinfo {author} {\bibfnamefont {K.}~\bibnamefont {Janssens}}, \bibinfo {author} {\bibfnamefont {J.}~\bibnamefont {Lawrence}}, \bibinfo {author} {\bibfnamefont {A.}~\bibnamefont {Effler}}, \bibinfo {author} {\bibfnamefont {R.~M.~S.}\ \bibnamefont {Schofield}}, \bibinfo {author} {\bibfnamefont {M.}~\bibnamefont {Lalleman}}, \bibinfo {author} {\bibfnamefont {J.}~\bibnamefont {Betzwieser}}, \bibinfo {author} {\bibfnamefont {N.}~\bibnamefont {Christensen}}, \bibinfo {author} {\bibfnamefont {M.~W.}\ \bibnamefont {Coughlin}}, \bibinfo {author} {\bibfnamefont {J.~C.}\ \bibnamefont {Driggers}}, \bibinfo {author} {\bibfnamefont {A.~F.}\ \bibnamefont {Helmling-Cornell}}, \bibinfo {author} {\bibfnamefont {T.~J.}\ \bibnamefont {O'Hanlon}}, \bibinfo {author} {\bibfnamefont {E.~A.}\ \bibnamefont {Quintero}}, \bibinfo {author} {\bibfnamefont {J.~A.~M.}\ \bibnamefont {Reis}},\ and\ \bibinfo {author} {\bibfnamefont {N.}~\bibnamefont {van Remortel}},\ }\href {https://arxiv.org/abs/2505.11903} {\bibinfo
  {title} {Coherent injection of magnetic noise and its impact on gravitational-wave searches}} (\bibinfo {year} {2025}),\ \Eprint {https://arxiv.org/abs/2505.11903} {arXiv:2505.11903 [gr-qc]} \BibitemShut {NoStop}%
\bibitem [{\citenamefont {Glanzer}\ \emph {et~al.}(2023{\natexlab{b}})\citenamefont {Glanzer}, \citenamefont {Banagiri}, \citenamefont {Coughlin}, \citenamefont {Soni}, \citenamefont {Zevin}, \citenamefont {Berry}, \citenamefont {Patane}, \citenamefont {Bahaadini}, \citenamefont {Rohani}, \citenamefont {Crowston}, \citenamefont {Kalogera}, \citenamefont {{\O}sterlund}, \citenamefont {Trouille},\ and\ \citenamefont {Katsaggelos}}]{Glanzer2023O3GSpyDQ}%
  \BibitemOpen
  \bibfield  {author} {\bibinfo {author} {\bibfnamefont {J.}~\bibnamefont {Glanzer}}, \bibinfo {author} {\bibfnamefont {S.}~\bibnamefont {Banagiri}}, \bibinfo {author} {\bibfnamefont {S.~B.}\ \bibnamefont {Coughlin}}, \bibinfo {author} {\bibfnamefont {S.}~\bibnamefont {Soni}}, \bibinfo {author} {\bibfnamefont {M.}~\bibnamefont {Zevin}}, \bibinfo {author} {\bibfnamefont {C.~P.~L.}\ \bibnamefont {Berry}}, \bibinfo {author} {\bibfnamefont {O.}~\bibnamefont {Patane}}, \bibinfo {author} {\bibfnamefont {S.}~\bibnamefont {Bahaadini}}, \bibinfo {author} {\bibfnamefont {N.}~\bibnamefont {Rohani}}, \bibinfo {author} {\bibfnamefont {K.}~\bibnamefont {Crowston}}, \bibinfo {author} {\bibfnamefont {V.}~\bibnamefont {Kalogera}}, \bibinfo {author} {\bibfnamefont {C.}~\bibnamefont {{\O}sterlund}}, \bibinfo {author} {\bibfnamefont {L.}~\bibnamefont {Trouille}},\ and\ \bibinfo {author} {\bibfnamefont {A.}~\bibnamefont {Katsaggelos}},\ }\href {https://doi.org/10.1088/1361-6382/acb633} {\bibfield  {journal} {\bibinfo  {journal}
  {Classical and Quantum Gravity}\ }\textbf {\bibinfo {volume} {40}},\ \bibinfo {pages} {065004} (\bibinfo {year} {2023}{\natexlab{b}})}\BibitemShut {NoStop}%
\bibitem [{\citenamefont {Helmling-Cornell}\ \emph {et~al.}(2024)\citenamefont {Helmling-Cornell}, \citenamefont {Nguyen}, \citenamefont {Schofield},\ and\ \citenamefont {Frey}}]{Helmling_Cornell2024autoenvcoupling}%
  \BibitemOpen
  \bibfield  {author} {\bibinfo {author} {\bibfnamefont {A.~F.}\ \bibnamefont {Helmling-Cornell}}, \bibinfo {author} {\bibfnamefont {P.}~\bibnamefont {Nguyen}}, \bibinfo {author} {\bibfnamefont {R.~M.~S.}\ \bibnamefont {Schofield}},\ and\ \bibinfo {author} {\bibfnamefont {R.}~\bibnamefont {Frey}},\ }\href {https://doi.org/10.1088/1361-6382/ad5139} {\bibfield  {journal} {\bibinfo  {journal} {Classical and Quantum Gravity}\ }\textbf {\bibinfo {volume} {41}},\ \bibinfo {pages} {145003} (\bibinfo {year} {2024})}\BibitemShut {NoStop}%
\bibitem [{\citenamefont {Smith}\ \emph {et~al.}(2011)\citenamefont {Smith}, \citenamefont {Abbott}, \citenamefont {Hirose}, \citenamefont {Leroy}, \citenamefont {MacLeod}, \citenamefont {McIver}, \citenamefont {Saulson},\ and\ \citenamefont {Shawhan}}]{Smith2011hveto}%
  \BibitemOpen
  \bibfield  {author} {\bibinfo {author} {\bibfnamefont {J.~R.}\ \bibnamefont {Smith}}, \bibinfo {author} {\bibfnamefont {T.}~\bibnamefont {Abbott}}, \bibinfo {author} {\bibfnamefont {E.}~\bibnamefont {Hirose}}, \bibinfo {author} {\bibfnamefont {N.}~\bibnamefont {Leroy}}, \bibinfo {author} {\bibfnamefont {D.}~\bibnamefont {MacLeod}}, \bibinfo {author} {\bibfnamefont {J.}~\bibnamefont {McIver}}, \bibinfo {author} {\bibfnamefont {P.}~\bibnamefont {Saulson}},\ and\ \bibinfo {author} {\bibfnamefont {P.}~\bibnamefont {Shawhan}},\ }\href {https://doi.org/10.1088/0264-9381/28/23/235005} {\bibfield  {journal} {\bibinfo  {journal} {Classical and Quantum Gravity}\ }\textbf {\bibinfo {volume} {28}},\ \bibinfo {pages} {235005} (\bibinfo {year} {2011})}\BibitemShut {NoStop}%
\bibitem [{\citenamefont {Huxford}\ \emph {et~al.}(2024)\citenamefont {Huxford}, \citenamefont {George}, \citenamefont {Trevor}, \citenamefont {Yarbrough},\ and\ \citenamefont {Godwin}}]{huxford2024performanceIDQ}%
  \BibitemOpen
  \bibfield  {author} {\bibinfo {author} {\bibfnamefont {R.}~\bibnamefont {Huxford}}, \bibinfo {author} {\bibfnamefont {R.}~\bibnamefont {George}}, \bibinfo {author} {\bibfnamefont {M.}~\bibnamefont {Trevor}}, \bibinfo {author} {\bibfnamefont {Z.}~\bibnamefont {Yarbrough}},\ and\ \bibinfo {author} {\bibfnamefont {P.}~\bibnamefont {Godwin}},\ }\href {https://arxiv.org/abs/2412.04638} {\bibinfo {title} {Performance of idq ahead of ligo, virgo, and kagra's fourth observing run}} (\bibinfo {year} {2024}),\ \Eprint {https://arxiv.org/abs/2412.04638} {arXiv:2412.04638 [gr-qc]} \BibitemShut {NoStop}%
\bibitem [{\citenamefont {Vazsonyi}\ and\ \citenamefont {Davis}(2023)}]{Vazsonyi2023Qtransform}%
  \BibitemOpen
  \bibfield  {author} {\bibinfo {author} {\bibfnamefont {L.}~\bibnamefont {Vazsonyi}}\ and\ \bibinfo {author} {\bibfnamefont {D.}~\bibnamefont {Davis}},\ }\href {https://doi.org/10.1088/1361-6382/acafd2} {\bibfield  {journal} {\bibinfo  {journal} {Classical and Quantum Gravity}\ }\textbf {\bibinfo {volume} {40}},\ \bibinfo {pages} {035008} (\bibinfo {year} {2023})}\BibitemShut {NoStop}%
\bibitem [{\citenamefont {Robinet}\ \emph {et~al.}(2020)\citenamefont {Robinet}, \citenamefont {Arnaud}, \citenamefont {Leroy}, \citenamefont {Lundgren}, \citenamefont {Macleod},\ and\ \citenamefont {McIver}}]{Robinet2020omicron}%
  \BibitemOpen
  \bibfield  {author} {\bibinfo {author} {\bibfnamefont {F.}~\bibnamefont {Robinet}}, \bibinfo {author} {\bibfnamefont {N.}~\bibnamefont {Arnaud}}, \bibinfo {author} {\bibfnamefont {N.}~\bibnamefont {Leroy}}, \bibinfo {author} {\bibfnamefont {A.}~\bibnamefont {Lundgren}}, \bibinfo {author} {\bibfnamefont {D.}~\bibnamefont {Macleod}},\ and\ \bibinfo {author} {\bibfnamefont {J.}~\bibnamefont {McIver}},\ }\href {https://doi.org/10.1016/j.softx.2020.100620} {\bibfield  {journal} {\bibinfo  {journal} {SoftwareX}\ }\textbf {\bibinfo {volume} {12}},\ \bibinfo {pages} {100620} (\bibinfo {year} {2020})}\BibitemShut {NoStop}%
\bibitem [{\citenamefont {Powell}\ \emph {et~al.}(2015)\citenamefont {Powell}, \citenamefont {Trifirò}, \citenamefont {Cuoco}, \citenamefont {Heng},\ and\ \citenamefont {Cavagli\a'}}]{Powell_2015}%
  \BibitemOpen
  \bibfield  {author} {\bibinfo {author} {\bibfnamefont {J.}~\bibnamefont {Powell}}, \bibinfo {author} {\bibfnamefont {D.}~\bibnamefont {Trifirò}}, \bibinfo {author} {\bibfnamefont {E.}~\bibnamefont {Cuoco}}, \bibinfo {author} {\bibfnamefont {I.~S.}\ \bibnamefont {Heng}},\ and\ \bibinfo {author} {\bibfnamefont {M.}~\bibnamefont {Cavagli\a'}},\ }\href {https://doi.org/10.1088/0264-9381/32/21/215012} {\bibfield  {journal} {\bibinfo  {journal} {Classical and Quantum Gravity}\ }\textbf {\bibinfo {volume} {32}},\ \bibinfo {pages} {215012} (\bibinfo {year} {2015})}\BibitemShut {NoStop}%
\bibitem [{\citenamefont {Macleod}\ \emph {et~al.}(2025)\citenamefont {Macleod}, \citenamefont {Goetz}, \citenamefont {Davis}, \citenamefont {Bidler}, \citenamefont {Smith}, \citenamefont {Lowry}, \citenamefont {Soni}, \citenamefont {Macedo}, \citenamefont {Lynam}, \citenamefont {Leman}, \citenamefont {Coughlin}, \citenamefont {Lundgren}, \citenamefont {Cotnoir},\ and\ \citenamefont {Massinger}}]{gwdetchar2025}%
  \BibitemOpen
  \bibfield  {author} {\bibinfo {author} {\bibfnamefont {D.}~\bibnamefont {Macleod}}, \bibinfo {author} {\bibfnamefont {E.}~\bibnamefont {Goetz}}, \bibinfo {author} {\bibfnamefont {D.}~\bibnamefont {Davis}}, \bibinfo {author} {\bibfnamefont {J.}~\bibnamefont {Bidler}}, \bibinfo {author} {\bibfnamefont {J.}~\bibnamefont {Smith}}, \bibinfo {author} {\bibfnamefont {M.}~\bibnamefont {Lowry}}, \bibinfo {author} {\bibfnamefont {S.}~\bibnamefont {Soni}}, \bibinfo {author} {\bibfnamefont {A.}~\bibnamefont {Macedo}}, \bibinfo {author} {\bibfnamefont {J.}~\bibnamefont {Lynam}}, \bibinfo {author} {\bibfnamefont {K.}~\bibnamefont {Leman}}, \bibinfo {author} {\bibfnamefont {S.}~\bibnamefont {Coughlin}}, \bibinfo {author} {\bibfnamefont {A.}~\bibnamefont {Lundgren}}, \bibinfo {author} {\bibfnamefont {L.}~\bibnamefont {Cotnoir}},\ and\ \bibinfo {author} {\bibfnamefont {T.}~\bibnamefont {Massinger}},\ }\href {https://doi.org/10.5281/zenodo.15530809} {\bibinfo {title} {gwdetchar/gwdetchar: 2.3.1}} (\bibinfo {year}
  {2025})\BibitemShut {NoStop}%
\bibitem [{\citenamefont {Zevin}\ \emph {et~al.}(2017)\citenamefont {Zevin}, \citenamefont {Coughlin}, \citenamefont {Bahaadini}, \citenamefont {Besler}, \citenamefont {Rohani}, \citenamefont {Allen}, \citenamefont {Cabero}, \citenamefont {Crowston}, \citenamefont {Katsaggelos}, \citenamefont {Larson}, \citenamefont {Lee}, \citenamefont {Lintott}, \citenamefont {Littenberg}, \citenamefont {Lundgren}, \citenamefont {{\o}sterlund}, \citenamefont {Smith}, \citenamefont {Trouille},\ and\ \citenamefont {Kalogera}}]{Zevin_2017GSpy}%
  \BibitemOpen
  \bibfield  {author} {\bibinfo {author} {\bibfnamefont {M.}~\bibnamefont {Zevin}}, \bibinfo {author} {\bibfnamefont {S.}~\bibnamefont {Coughlin}}, \bibinfo {author} {\bibfnamefont {S.}~\bibnamefont {Bahaadini}}, \bibinfo {author} {\bibfnamefont {E.}~\bibnamefont {Besler}}, \bibinfo {author} {\bibfnamefont {N.}~\bibnamefont {Rohani}}, \bibinfo {author} {\bibfnamefont {S.}~\bibnamefont {Allen}}, \bibinfo {author} {\bibfnamefont {M.}~\bibnamefont {Cabero}}, \bibinfo {author} {\bibfnamefont {K.}~\bibnamefont {Crowston}}, \bibinfo {author} {\bibfnamefont {A.~K.}\ \bibnamefont {Katsaggelos}}, \bibinfo {author} {\bibfnamefont {S.~L.}\ \bibnamefont {Larson}}, \bibinfo {author} {\bibfnamefont {T.~K.}\ \bibnamefont {Lee}}, \bibinfo {author} {\bibfnamefont {C.}~\bibnamefont {Lintott}}, \bibinfo {author} {\bibfnamefont {T.~B.}\ \bibnamefont {Littenberg}}, \bibinfo {author} {\bibfnamefont {A.}~\bibnamefont {Lundgren}}, \bibinfo {author} {\bibfnamefont {C.}~\bibnamefont {{\o}sterlund}}, \bibinfo {author} {\bibfnamefont
  {J.~R.}\ \bibnamefont {Smith}}, \bibinfo {author} {\bibfnamefont {L.}~\bibnamefont {Trouille}},\ and\ \bibinfo {author} {\bibfnamefont {V.}~\bibnamefont {Kalogera}},\ }\href {https://doi.org/10.1088/1361-6382/aa5cea} {\bibfield  {journal} {\bibinfo  {journal} {Classical and Quantum Gravity}\ }\textbf {\bibinfo {volume} {34}},\ \bibinfo {pages} {064003} (\bibinfo {year} {2017})}\BibitemShut {NoStop}%
\bibitem [{\citenamefont {Coughlin}\ \emph {et~al.}(2019)\citenamefont {Coughlin} \emph {et~al.}}]{Coughlin:2019ref}%
  \BibitemOpen
  \bibfield  {author} {\bibinfo {author} {\bibfnamefont {S.~B.}\ \bibnamefont {Coughlin}} \emph {et~al.},\ }\href {https://doi.org/10.1103/PhysRevD.99.082002} {\bibfield  {journal} {\bibinfo  {journal} {Phys. Rev. D}\ }\textbf {\bibinfo {volume} {99}},\ \bibinfo {pages} {082002} (\bibinfo {year} {2019})},\ \Eprint {https://arxiv.org/abs/1903.04058} {arXiv:1903.04058 [astro-ph.IM]} \BibitemShut {NoStop}%
\bibitem [{\citenamefont {Soni}\ \emph {et~al.}(2021)\citenamefont {Soni}, \citenamefont {Berry}, \citenamefont {Coughlin}, \citenamefont {Harandi}, \citenamefont {Jackson}, \citenamefont {Crowston}, \citenamefont {{\O}sterlund}, \citenamefont {Patane}, \citenamefont {Katsaggelos}, \citenamefont {Trouille} \emph {et~al.}}]{soni2021discovering}%
  \BibitemOpen
  \bibfield  {author} {\bibinfo {author} {\bibfnamefont {S.}~\bibnamefont {Soni}}, \bibinfo {author} {\bibfnamefont {C.~P.~L.}\ \bibnamefont {Berry}}, \bibinfo {author} {\bibfnamefont {S.~B.}\ \bibnamefont {Coughlin}}, \bibinfo {author} {\bibfnamefont {M.}~\bibnamefont {Harandi}}, \bibinfo {author} {\bibfnamefont {C.~B.}\ \bibnamefont {Jackson}}, \bibinfo {author} {\bibfnamefont {K.}~\bibnamefont {Crowston}}, \bibinfo {author} {\bibfnamefont {C.}~\bibnamefont {{\O}sterlund}}, \bibinfo {author} {\bibfnamefont {O.}~\bibnamefont {Patane}}, \bibinfo {author} {\bibfnamefont {A.~K.}\ \bibnamefont {Katsaggelos}}, \bibinfo {author} {\bibfnamefont {L.}~\bibnamefont {Trouille}}, \emph {et~al.},\ }\href@noop {} {\bibfield  {journal} {\bibinfo  {journal} {Classical and Quantum Gravity}\ }\textbf {\bibinfo {volume} {38}},\ \bibinfo {pages} {195016} (\bibinfo {year} {2021})}\BibitemShut {NoStop}%
\bibitem [{\citenamefont {Davis}\ \emph {et~al.}(2019)\citenamefont {Davis}, \citenamefont {Massinger}, \citenamefont {Lundgren}, \citenamefont {Driggers}, \citenamefont {Urban},\ and\ \citenamefont {Nuttall}}]{Davis2019improvingSens}%
  \BibitemOpen
  \bibfield  {author} {\bibinfo {author} {\bibfnamefont {D.}~\bibnamefont {Davis}}, \bibinfo {author} {\bibfnamefont {T.}~\bibnamefont {Massinger}}, \bibinfo {author} {\bibfnamefont {A.}~\bibnamefont {Lundgren}}, \bibinfo {author} {\bibfnamefont {J.~C.}\ \bibnamefont {Driggers}}, \bibinfo {author} {\bibfnamefont {A.~L.}\ \bibnamefont {Urban}},\ and\ \bibinfo {author} {\bibfnamefont {L.}~\bibnamefont {Nuttall}},\ }\href {https://doi.org/10.1088/1361-6382/ab01c5} {\bibfield  {journal} {\bibinfo  {journal} {Classical and Quantum Gravity}\ }\textbf {\bibinfo {volume} {36}},\ \bibinfo {pages} {055011} (\bibinfo {year} {2019})}\BibitemShut {NoStop}%
\bibitem [{\citenamefont {Pankow}\ \emph {et~al.}(2018{\natexlab{b}})\citenamefont {Pankow} \emph {et~al.}}]{Pankow:2018qpo}%
  \BibitemOpen
  \bibfield  {author} {\bibinfo {author} {\bibfnamefont {C.}~\bibnamefont {Pankow}} \emph {et~al.},\ }\href {https://doi.org/10.1103/PhysRevD.98.084016} {\bibfield  {journal} {\bibinfo  {journal} {Phys. Rev. D}\ }\textbf {\bibinfo {volume} {98}},\ \bibinfo {pages} {084016} (\bibinfo {year} {2018}{\natexlab{b}})},\ \Eprint {https://arxiv.org/abs/1808.03619} {arXiv:1808.03619 [gr-qc]} \BibitemShut {NoStop}%
\bibitem [{\citenamefont {Davis}\ \emph {et~al.}(2022)\citenamefont {Davis}, \citenamefont {Littenberg}, \citenamefont {Romero-Shaw}, \citenamefont {Millhouse}, \citenamefont {McIver}, \citenamefont {Di~Renzo},\ and\ \citenamefont {Ashton}}]{Davis2022subtractingGlitches}%
  \BibitemOpen
  \bibfield  {author} {\bibinfo {author} {\bibfnamefont {D.}~\bibnamefont {Davis}}, \bibinfo {author} {\bibfnamefont {T.~B.}\ \bibnamefont {Littenberg}}, \bibinfo {author} {\bibfnamefont {I.~M.}\ \bibnamefont {Romero-Shaw}}, \bibinfo {author} {\bibfnamefont {M.}~\bibnamefont {Millhouse}}, \bibinfo {author} {\bibfnamefont {J.}~\bibnamefont {McIver}}, \bibinfo {author} {\bibfnamefont {F.}~\bibnamefont {Di~Renzo}},\ and\ \bibinfo {author} {\bibfnamefont {G.}~\bibnamefont {Ashton}},\ }\href {https://doi.org/10.1088/1361-6382/aca238} {\bibfield  {journal} {\bibinfo  {journal} {Classical and Quantum Gravity}\ }\textbf {\bibinfo {volume} {39}},\ \bibinfo {pages} {245013} (\bibinfo {year} {2022})}\BibitemShut {NoStop}%
\bibitem [{\citenamefont {Davis}\ \emph {et~al.}(2021{\natexlab{a}})\citenamefont {Davis} \emph {et~al.}}]{LIGO:2021ppb}%
  \BibitemOpen
  \bibfield  {author} {\bibinfo {author} {\bibfnamefont {D.}~\bibnamefont {Davis}} \emph {et~al.} (\bibinfo {collaboration} {LIGO}),\ }\href {https://doi.org/10.1088/1361-6382/abfd85} {\bibfield  {journal} {\bibinfo  {journal} {Class. Quant. Grav.}\ }\textbf {\bibinfo {volume} {38}},\ \bibinfo {pages} {135014} (\bibinfo {year} {2021}{\natexlab{a}})},\ \Eprint {https://arxiv.org/abs/2101.11673} {arXiv:2101.11673 [astro-ph.IM]} \BibitemShut {NoStop}%
\bibitem [{\citenamefont {Hanna}(2006)}]{Hanna2006falsealarm}%
  \BibitemOpen
  \bibfield  {author} {\bibinfo {author} {\bibfnamefont {C.~R.}\ \bibnamefont {Hanna}},\ }\href {https://doi.org/10.1088/0264-9381/23/8/S03} {\bibfield  {journal} {\bibinfo  {journal} {Classical and Quantum Gravity}\ }\textbf {\bibinfo {volume} {23}},\ \bibinfo {pages} {S17} (\bibinfo {year} {2006})}\BibitemShut {NoStop}%
\bibitem [{\citenamefont {Essick}\ \emph {et~al.}(2020)\citenamefont {Essick}, \citenamefont {Godwin}, \citenamefont {Hanna}, \citenamefont {Blackburn},\ and\ \citenamefont {Katsavounidis}}]{essick2020idq}%
  \BibitemOpen
  \bibfield  {author} {\bibinfo {author} {\bibfnamefont {R.}~\bibnamefont {Essick}}, \bibinfo {author} {\bibfnamefont {P.}~\bibnamefont {Godwin}}, \bibinfo {author} {\bibfnamefont {C.}~\bibnamefont {Hanna}}, \bibinfo {author} {\bibfnamefont {L.}~\bibnamefont {Blackburn}},\ and\ \bibinfo {author} {\bibfnamefont {E.}~\bibnamefont {Katsavounidis}},\ }\href {https://arxiv.org/abs/2005.12761} {\bibinfo {title} {idq: Statistical inference of non-gaussian noise with auxiliary degrees of freedom in gravitational-wave detectors}} (\bibinfo {year} {2020}),\ \Eprint {https://arxiv.org/abs/2005.12761} {arXiv:2005.12761 [astro-ph.IM]} \BibitemShut {NoStop}%
\bibitem [{\citenamefont {Zevin}\ \emph {et~al.}(2024)\citenamefont {Zevin}, \citenamefont {Jackson}, \citenamefont {Doctor}, \citenamefont {Wu}, \citenamefont {{\o}sterlund}, \citenamefont {Johnson}, \citenamefont {Berry}, \citenamefont {Crowston}, \citenamefont {Coughlin}, \citenamefont {Kalogera}, \citenamefont {Banagiri}, \citenamefont {Davis}, \citenamefont {Glanzer}, \citenamefont {Hao}, \citenamefont {Katsaggelos}, \citenamefont {Patane}, \citenamefont {Sanchez}, \citenamefont {Smith}, \citenamefont {Soni}, \citenamefont {Trouille}, \citenamefont {Walker}, \citenamefont {Aerith}, \citenamefont {Domainko}, \citenamefont {Baranowski}, \citenamefont {Niklasch},\ and\ \citenamefont {T\'egl\'as}}]{Zevin_2024}%
  \BibitemOpen
  \bibfield  {author} {\bibinfo {author} {\bibfnamefont {M.}~\bibnamefont {Zevin}}, \bibinfo {author} {\bibfnamefont {C.~B.}\ \bibnamefont {Jackson}}, \bibinfo {author} {\bibfnamefont {Z.}~\bibnamefont {Doctor}}, \bibinfo {author} {\bibfnamefont {Y.}~\bibnamefont {Wu}}, \bibinfo {author} {\bibfnamefont {C.}~\bibnamefont {{\o}sterlund}}, \bibinfo {author} {\bibfnamefont {L.~C.}\ \bibnamefont {Johnson}}, \bibinfo {author} {\bibfnamefont {C.~P.~L.}\ \bibnamefont {Berry}}, \bibinfo {author} {\bibfnamefont {K.}~\bibnamefont {Crowston}}, \bibinfo {author} {\bibfnamefont {S.~B.}\ \bibnamefont {Coughlin}}, \bibinfo {author} {\bibfnamefont {V.}~\bibnamefont {Kalogera}}, \bibinfo {author} {\bibfnamefont {S.}~\bibnamefont {Banagiri}}, \bibinfo {author} {\bibfnamefont {D.}~\bibnamefont {Davis}}, \bibinfo {author} {\bibfnamefont {J.}~\bibnamefont {Glanzer}}, \bibinfo {author} {\bibfnamefont {R.}~\bibnamefont {Hao}}, \bibinfo {author} {\bibfnamefont {A.~K.}\ \bibnamefont {Katsaggelos}}, \bibinfo {author} {\bibfnamefont
  {O.}~\bibnamefont {Patane}}, \bibinfo {author} {\bibfnamefont {J.}~\bibnamefont {Sanchez}}, \bibinfo {author} {\bibfnamefont {J.}~\bibnamefont {Smith}}, \bibinfo {author} {\bibfnamefont {S.}~\bibnamefont {Soni}}, \bibinfo {author} {\bibfnamefont {L.}~\bibnamefont {Trouille}}, \bibinfo {author} {\bibfnamefont {M.}~\bibnamefont {Walker}}, \bibinfo {author} {\bibfnamefont {I.}~\bibnamefont {Aerith}}, \bibinfo {author} {\bibfnamefont {W.}~\bibnamefont {Domainko}}, \bibinfo {author} {\bibfnamefont {V.-G.}\ \bibnamefont {Baranowski}}, \bibinfo {author} {\bibfnamefont {G.}~\bibnamefont {Niklasch}},\ and\ \bibinfo {author} {\bibfnamefont {B.}~\bibnamefont {T\'egl\'as}},\ }\bibfield  {journal} {\bibinfo  {journal} {The European Physical Journal Plus}\ }\textbf {\bibinfo {volume} {139}},\ \href {https://doi.org/10.1140/epjp/s13360-023-04795-4} {10.1140/epjp/s13360-023-04795-4} (\bibinfo {year} {2024})\BibitemShut {NoStop}%
\bibitem [{\citenamefont {George}\ \emph {et~al.}(2018)\citenamefont {George}, \citenamefont {Shen},\ and\ \citenamefont {Huerta}}]{George2018DeepTransferLearning}%
  \BibitemOpen
  \bibfield  {author} {\bibinfo {author} {\bibfnamefont {D.}~\bibnamefont {George}}, \bibinfo {author} {\bibfnamefont {H.}~\bibnamefont {Shen}},\ and\ \bibinfo {author} {\bibfnamefont {E.}~\bibnamefont {Huerta}},\ }\bibfield  {journal} {\bibinfo  {journal} {Physical Review D}\ }\textbf {\bibinfo {volume} {97}},\ \href {https://doi.org/10.1103/physrevd.97.101501} {10.1103/physrevd.97.101501} (\bibinfo {year} {2018})\BibitemShut {NoStop}%
\bibitem [{\citenamefont {Bahaadini}\ \emph {et~al.}(2018)\citenamefont {Bahaadini}, \citenamefont {Noroozi}, \citenamefont {Rohani}, \citenamefont {Coughlin}, \citenamefont {Zevin},\ and\ \citenamefont {Katsaggelos}}]{bahaadini2018DIRECT}%
  \BibitemOpen
  \bibfield  {author} {\bibinfo {author} {\bibfnamefont {S.}~\bibnamefont {Bahaadini}}, \bibinfo {author} {\bibfnamefont {V.}~\bibnamefont {Noroozi}}, \bibinfo {author} {\bibfnamefont {N.}~\bibnamefont {Rohani}}, \bibinfo {author} {\bibfnamefont {S.}~\bibnamefont {Coughlin}}, \bibinfo {author} {\bibfnamefont {M.}~\bibnamefont {Zevin}},\ and\ \bibinfo {author} {\bibfnamefont {A.~K.}\ \bibnamefont {Katsaggelos}},\ }\href {http://arxiv.org/abs/1805.02296} {\bibfield  {journal} {\bibinfo  {journal} {CoRR}\ }\textbf {\bibinfo {volume} {abs/1805.02296}} (\bibinfo {year} {2018})},\ \Eprint {https://arxiv.org/abs/1805.02296} {1805.02296} \BibitemShut {NoStop}%
\bibitem [{\citenamefont {Cuoco}\ \emph {et~al.}(2020)\citenamefont {Cuoco}, \citenamefont {Powell}, \citenamefont {Cavagli\a'}, \citenamefont {Ackley}, \citenamefont {Bejger}, \citenamefont {Chatterjee}, \citenamefont {Coughlin}, \citenamefont {Coughlin}, \citenamefont {Easter}, \citenamefont {Essick}, \citenamefont {Gabbard}, \citenamefont {Gebhard}, \citenamefont {Ghosh}, \citenamefont {Haegel}, \citenamefont {Iess}, \citenamefont {Keitel}, \citenamefont {M\'arka}, \citenamefont {M\'arka}, \citenamefont {Morawski}, \citenamefont {Nguyen}, \citenamefont {Ormiston}, \citenamefont {P\"urrer}, \citenamefont {Razzano}, \citenamefont {Staats}, \citenamefont {Vajente},\ and\ \citenamefont {Williams}}]{Cuoco2020enhanceGWML}%
  \BibitemOpen
  \bibfield  {author} {\bibinfo {author} {\bibfnamefont {E.}~\bibnamefont {Cuoco}}, \bibinfo {author} {\bibfnamefont {J.}~\bibnamefont {Powell}}, \bibinfo {author} {\bibfnamefont {M.}~\bibnamefont {Cavagli\a'}}, \bibinfo {author} {\bibfnamefont {K.}~\bibnamefont {Ackley}}, \bibinfo {author} {\bibfnamefont {M.}~\bibnamefont {Bejger}}, \bibinfo {author} {\bibfnamefont {C.}~\bibnamefont {Chatterjee}}, \bibinfo {author} {\bibfnamefont {M.}~\bibnamefont {Coughlin}}, \bibinfo {author} {\bibfnamefont {S.}~\bibnamefont {Coughlin}}, \bibinfo {author} {\bibfnamefont {P.}~\bibnamefont {Easter}}, \bibinfo {author} {\bibfnamefont {R.}~\bibnamefont {Essick}}, \bibinfo {author} {\bibfnamefont {H.}~\bibnamefont {Gabbard}}, \bibinfo {author} {\bibfnamefont {T.}~\bibnamefont {Gebhard}}, \bibinfo {author} {\bibfnamefont {S.}~\bibnamefont {Ghosh}}, \bibinfo {author} {\bibfnamefont {L.}~\bibnamefont {Haegel}}, \bibinfo {author} {\bibfnamefont {A.}~\bibnamefont {Iess}}, \bibinfo {author} {\bibfnamefont {D.}~\bibnamefont {Keitel}},
  \bibinfo {author} {\bibfnamefont {Z.}~\bibnamefont {M\'arka}}, \bibinfo {author} {\bibfnamefont {S.}~\bibnamefont {M\'arka}}, \bibinfo {author} {\bibfnamefont {F.}~\bibnamefont {Morawski}}, \bibinfo {author} {\bibfnamefont {T.}~\bibnamefont {Nguyen}}, \bibinfo {author} {\bibfnamefont {R.}~\bibnamefont {Ormiston}}, \bibinfo {author} {\bibfnamefont {M.}~\bibnamefont {P\"urrer}}, \bibinfo {author} {\bibfnamefont {M.}~\bibnamefont {Razzano}}, \bibinfo {author} {\bibfnamefont {K.}~\bibnamefont {Staats}}, \bibinfo {author} {\bibfnamefont {G.}~\bibnamefont {Vajente}},\ and\ \bibinfo {author} {\bibfnamefont {D.}~\bibnamefont {Williams}},\ }\href {https://doi.org/10.1088/2632-2153/abb93a} {\bibfield  {journal} {\bibinfo  {journal} {Machine Learning: Science and Technology}\ }\textbf {\bibinfo {volume} {2}},\ \bibinfo {pages} {011002} (\bibinfo {year} {2020})}\BibitemShut {NoStop}%
\bibitem [{\citenamefont {Jung}\ \emph {et~al.}(2022)\citenamefont {Jung}, \citenamefont {Oh}, \citenamefont {Kim}, \citenamefont {Son}, \citenamefont {Yokozawa}, \citenamefont {Washimi},\ and\ \citenamefont {Oh}}]{Jung2022channelCouplings}%
  \BibitemOpen
  \bibfield  {author} {\bibinfo {author} {\bibfnamefont {P.}~\bibnamefont {Jung}}, \bibinfo {author} {\bibfnamefont {S.~H.}\ \bibnamefont {Oh}}, \bibinfo {author} {\bibfnamefont {Y.-M.}\ \bibnamefont {Kim}}, \bibinfo {author} {\bibfnamefont {E.~J.}\ \bibnamefont {Son}}, \bibinfo {author} {\bibfnamefont {T.}~\bibnamefont {Yokozawa}}, \bibinfo {author} {\bibfnamefont {T.}~\bibnamefont {Washimi}},\ and\ \bibinfo {author} {\bibfnamefont {J.~J.}\ \bibnamefont {Oh}},\ }\bibfield  {journal} {\bibinfo  {journal} {Physical Review D}\ }\textbf {\bibinfo {volume} {106}},\ \href {https://doi.org/10.1103/physrevd.106.042010} {10.1103/physrevd.106.042010} (\bibinfo {year} {2022})\BibitemShut {NoStop}%
\bibitem [{\citenamefont {Essick}\ \emph {et~al.}(2013)\citenamefont {Essick}, \citenamefont {Blackburn},\ and\ \citenamefont {Katsavounidis}}]{Essick2013optVetoes}%
  \BibitemOpen
  \bibfield  {author} {\bibinfo {author} {\bibfnamefont {R.}~\bibnamefont {Essick}}, \bibinfo {author} {\bibfnamefont {L.}~\bibnamefont {Blackburn}},\ and\ \bibinfo {author} {\bibfnamefont {E.}~\bibnamefont {Katsavounidis}},\ }\href {https://doi.org/10.1088/0264-9381/30/15/155010} {\bibfield  {journal} {\bibinfo  {journal} {Classical and Quantum Gravity}\ }\textbf {\bibinfo {volume} {30}},\ \bibinfo {pages} {155010} (\bibinfo {year} {2013})}\BibitemShut {NoStop}%
\bibitem [{\citenamefont {Fiori}\ \emph {et~al.}(2020)\citenamefont {Fiori}, \citenamefont {Effler}, \citenamefont {Nguyen}, \citenamefont {Paoletti}, \citenamefont {Schofield},\ and\ \citenamefont {Tringali}}]{Fiori2020envNoiseBook}%
  \BibitemOpen
  \bibfield  {author} {\bibinfo {author} {\bibfnamefont {I.}~\bibnamefont {Fiori}}, \bibinfo {author} {\bibfnamefont {A.}~\bibnamefont {Effler}}, \bibinfo {author} {\bibfnamefont {P.}~\bibnamefont {Nguyen}}, \bibinfo {author} {\bibfnamefont {F.}~\bibnamefont {Paoletti}}, \bibinfo {author} {\bibfnamefont {R.~M.~S.}\ \bibnamefont {Schofield}},\ and\ \bibinfo {author} {\bibfnamefont {M.~C.}\ \bibnamefont {Tringali}},\ }\bibinfo {title} {Environmental noise in gravitational-wave interferometers},\ in\ \href {https://doi.org/10.1007/978-981-15-4702-7_10-1} {\emph {\bibinfo {booktitle} {Handbook of Gravitational Wave Astronomy}}},\ \bibinfo {editor} {edited by\ \bibinfo {editor} {\bibfnamefont {C.}~\bibnamefont {Bambi}}, \bibinfo {editor} {\bibfnamefont {S.}~\bibnamefont {Katsanevas}},\ and\ \bibinfo {editor} {\bibfnamefont {K.~D.}\ \bibnamefont {Kokkotas}}}\ (\bibinfo  {publisher} {Springer Singapore},\ \bibinfo {address} {Singapore},\ \bibinfo {year} {2020})\ pp.\ \bibinfo {pages} {1--72}\BibitemShut {NoStop}%
\bibitem [{\citenamefont {Van~der Maaten}\ and\ \citenamefont {Hinton}(2008)}]{vander2008visualizing}%
  \BibitemOpen
  \bibfield  {author} {\bibinfo {author} {\bibfnamefont {L.}~\bibnamefont {Van~der Maaten}}\ and\ \bibinfo {author} {\bibfnamefont {G.}~\bibnamefont {Hinton}},\ }\href@noop {} {\bibfield  {journal} {\bibinfo  {journal} {Journal of machine learning research}\ }\textbf {\bibinfo {volume} {9}} (\bibinfo {year} {2008})}\BibitemShut {NoStop}%
\bibitem [{\citenamefont {Ajith}\ \emph {et~al.}(2014)\citenamefont {Ajith}, \citenamefont {Isogai}, \citenamefont {Christensen}, \citenamefont {Adhikari}, \citenamefont {Pearlman}, \citenamefont {Wein}, \citenamefont {Weinstein},\ and\ \citenamefont {Yuan}}]{Ajith2014bilinearcoupling}%
  \BibitemOpen
  \bibfield  {author} {\bibinfo {author} {\bibfnamefont {P.}~\bibnamefont {Ajith}}, \bibinfo {author} {\bibfnamefont {T.}~\bibnamefont {Isogai}}, \bibinfo {author} {\bibfnamefont {N.}~\bibnamefont {Christensen}}, \bibinfo {author} {\bibfnamefont {R.~X.}\ \bibnamefont {Adhikari}}, \bibinfo {author} {\bibfnamefont {A.~B.}\ \bibnamefont {Pearlman}}, \bibinfo {author} {\bibfnamefont {A.}~\bibnamefont {Wein}}, \bibinfo {author} {\bibfnamefont {A.~J.}\ \bibnamefont {Weinstein}},\ and\ \bibinfo {author} {\bibfnamefont {B.}~\bibnamefont {Yuan}},\ }\bibfield  {journal} {\bibinfo  {journal} {Physical Review D}\ }\textbf {\bibinfo {volume} {89}},\ \href {https://doi.org/10.1103/physrevd.89.122001} {10.1103/physrevd.89.122001} (\bibinfo {year} {2014})\BibitemShut {NoStop}%
\bibitem [{\citenamefont {Bose}\ \emph {et~al.}(2016)\citenamefont {Bose}, \citenamefont {Hall}, \citenamefont {Mazumder}, \citenamefont {Dhurandhar}, \citenamefont {Gupta},\ and\ \citenamefont {Lundgren}}]{Bose2016BiNonlinearCoupling}%
  \BibitemOpen
  \bibfield  {author} {\bibinfo {author} {\bibfnamefont {S.}~\bibnamefont {Bose}}, \bibinfo {author} {\bibfnamefont {B.}~\bibnamefont {Hall}}, \bibinfo {author} {\bibfnamefont {N.}~\bibnamefont {Mazumder}}, \bibinfo {author} {\bibfnamefont {S.}~\bibnamefont {Dhurandhar}}, \bibinfo {author} {\bibfnamefont {A.}~\bibnamefont {Gupta}},\ and\ \bibinfo {author} {\bibfnamefont {A.}~\bibnamefont {Lundgren}},\ }\href {https://doi.org/10.1088/1742-6596/716/1/012007} {\bibfield  {journal} {\bibinfo  {journal} {Journal of Physics: Conference Series}\ }\textbf {\bibinfo {volume} {716}},\ \bibinfo {pages} {012007} (\bibinfo {year} {2016})}\BibitemShut {NoStop}%
\bibitem [{\citenamefont {Davis}\ \emph {et~al.}(2021{\natexlab{b}})\citenamefont {Davis} \emph {et~al.}}]{Davis2021detcharO2O3}%
  \BibitemOpen
  \bibfield  {author} {\bibinfo {author} {\bibfnamefont {D.}~\bibnamefont {Davis}} \emph {et~al.},\ }\href {https://doi.org/10.1088/1361-6382/abfd85} {\bibfield  {journal} {\bibinfo  {journal} {Classical and Quantum Gravity}\ }\textbf {\bibinfo {volume} {38}},\ \bibinfo {pages} {135014} (\bibinfo {year} {2021}{\natexlab{b}})}\BibitemShut {NoStop}%
\bibitem [{\citenamefont {Soni}\ \emph {et~al.}(2020)\citenamefont {Soni}, \citenamefont {Austin}, \citenamefont {Effler}, \citenamefont {Schofield}, \citenamefont {Gonz{\'a}lez}, \citenamefont {Frolov}, \citenamefont {Driggers}, \citenamefont {Pele}, \citenamefont {Urban}, \citenamefont {Valdes} \emph {et~al.}}]{soni2020scatter}%
  \BibitemOpen
  \bibfield  {author} {\bibinfo {author} {\bibfnamefont {S.}~\bibnamefont {Soni}}, \bibinfo {author} {\bibfnamefont {C.}~\bibnamefont {Austin}}, \bibinfo {author} {\bibfnamefont {A.}~\bibnamefont {Effler}}, \bibinfo {author} {\bibfnamefont {R.}~\bibnamefont {Schofield}}, \bibinfo {author} {\bibfnamefont {G.}~\bibnamefont {Gonz{\'a}lez}}, \bibinfo {author} {\bibfnamefont {V.}~\bibnamefont {Frolov}}, \bibinfo {author} {\bibfnamefont {J.~C.}\ \bibnamefont {Driggers}}, \bibinfo {author} {\bibfnamefont {A.}~\bibnamefont {Pele}}, \bibinfo {author} {\bibfnamefont {A.}~\bibnamefont {Urban}}, \bibinfo {author} {\bibfnamefont {G.}~\bibnamefont {Valdes}}, \emph {et~al.},\ }\href@noop {} {\bibfield  {journal} {\bibinfo  {journal} {Classical and Quantum Gravity}\ }\textbf {\bibinfo {volume} {38}},\ \bibinfo {pages} {025016} (\bibinfo {year} {2020})}\BibitemShut {NoStop}%
\bibitem [{\citenamefont {Udall}\ and\ \citenamefont {Davis}(2023)}]{Udall2023bayesianscatter}%
  \BibitemOpen
  \bibfield  {author} {\bibinfo {author} {\bibfnamefont {R.~P.}\ \bibnamefont {Udall}}\ and\ \bibinfo {author} {\bibfnamefont {D.}~\bibnamefont {Davis}},\ }\bibfield  {journal} {\bibinfo  {journal} {Applied Physics Letters}\ }\textbf {\bibinfo {volume} {122}},\ \href {https://doi.org/10.1063/5.0136896} {10.1063/5.0136896} (\bibinfo {year} {2023})\BibitemShut {NoStop}%
\bibitem [{\citenamefont {Vinet}\ \emph {et~al.}(1996)\citenamefont {Vinet}, \citenamefont {Brisson},\ and\ \citenamefont {Braccini}}]{vinet1996scatteredlight}%
  \BibitemOpen
  \bibfield  {author} {\bibinfo {author} {\bibfnamefont {J.-Y.}\ \bibnamefont {Vinet}}, \bibinfo {author} {\bibfnamefont {V.}~\bibnamefont {Brisson}},\ and\ \bibinfo {author} {\bibfnamefont {S.}~\bibnamefont {Braccini}},\ }\href {https://doi.org/10.1103/PhysRevD.54.1276} {\bibfield  {journal} {\bibinfo  {journal} {Physical Review D}\ }\textbf {\bibinfo {volume} {54}},\ \bibinfo {pages} {1276} (\bibinfo {year} {1996})}\BibitemShut {NoStop}%
\bibitem [{\citenamefont {Mackenzie}\ \emph {et~al.}(2025)\citenamefont {Mackenzie}, \citenamefont {Berry}, \citenamefont {Niklasch}, \citenamefont {T\'egl\'as}, \citenamefont {Unsworth}, \citenamefont {Crowston}, \citenamefont {Davis},\ and\ \citenamefont {Katsaggelos}}]{mackenzie2025huntingnewglitches}%
  \BibitemOpen
  \bibfield  {author} {\bibinfo {author} {\bibfnamefont {E.}~\bibnamefont {Mackenzie}}, \bibinfo {author} {\bibfnamefont {C.~P.~L.}\ \bibnamefont {Berry}}, \bibinfo {author} {\bibfnamefont {G.}~\bibnamefont {Niklasch}}, \bibinfo {author} {\bibfnamefont {B.}~\bibnamefont {T\'egl\'as}}, \bibinfo {author} {\bibfnamefont {C.}~\bibnamefont {Unsworth}}, \bibinfo {author} {\bibfnamefont {K.}~\bibnamefont {Crowston}}, \bibinfo {author} {\bibfnamefont {D.}~\bibnamefont {Davis}},\ and\ \bibinfo {author} {\bibfnamefont {A.~K.}\ \bibnamefont {Katsaggelos}},\ }\href {https://arxiv.org/abs/2508.13923} {\bibinfo {title} {Hunting for new glitches in ligo data using community science}} (\bibinfo {year} {2025}),\ \Eprint {https://arxiv.org/abs/2508.13923} {arXiv:2508.13923 [gr-qc]} \BibitemShut {NoStop}%
\bibitem [{\citenamefont {Garaventa}\ \emph {et~al.}(2024)\citenamefont {Garaventa} \emph {et~al.}}]{Garaventa02024nextgen}%
  \BibitemOpen
  \bibfield  {author} {\bibinfo {author} {\bibfnamefont {B.}~\bibnamefont {Garaventa}} \emph {et~al.},\ }\href {https://doi.org/10.1016/j.nima.2024.169629} {\bibfield  {journal} {\bibinfo  {journal} {Nucl. Instrum. Meth. A}\ }\textbf {\bibinfo {volume} {1066}},\ \bibinfo {pages} {169629} (\bibinfo {year} {2024})}\BibitemShut {NoStop}%
\bibitem [{\citenamefont {Capote}\ \emph {et~al.}(2024)\citenamefont {Capote}, \citenamefont {Dartez},\ and\ \citenamefont {Davis}}]{capote2024DQfornextgen}%
  \BibitemOpen
  \bibfield  {author} {\bibinfo {author} {\bibfnamefont {E.}~\bibnamefont {Capote}}, \bibinfo {author} {\bibfnamefont {L.}~\bibnamefont {Dartez}},\ and\ \bibinfo {author} {\bibfnamefont {D.}~\bibnamefont {Davis}},\ }\href {https://arxiv.org/abs/2404.04761} {\bibinfo {title} {Technical noise, data quality, and calibration requirements for next-generation gravitational-wave science}} (\bibinfo {year} {2024}),\ \Eprint {https://arxiv.org/abs/2404.04761} {arXiv:2404.04761 [astro-ph.IM]} \BibitemShut {NoStop}%
\bibitem [{\citenamefont {Srivastava}\ \emph {et~al.}(2022)\citenamefont {Srivastava}, \citenamefont {Davis}, \citenamefont {Kuns}, \citenamefont {Landry}, \citenamefont {Ballmer}, \citenamefont {Evans}, \citenamefont {Hall}, \citenamefont {Read},\ and\ \citenamefont {Sathyaprakash}}]{Srivastava2022CEdetectors}%
  \BibitemOpen
  \bibfield  {author} {\bibinfo {author} {\bibfnamefont {V.}~\bibnamefont {Srivastava}}, \bibinfo {author} {\bibfnamefont {D.}~\bibnamefont {Davis}}, \bibinfo {author} {\bibfnamefont {K.}~\bibnamefont {Kuns}}, \bibinfo {author} {\bibfnamefont {P.}~\bibnamefont {Landry}}, \bibinfo {author} {\bibfnamefont {S.}~\bibnamefont {Ballmer}}, \bibinfo {author} {\bibfnamefont {M.}~\bibnamefont {Evans}}, \bibinfo {author} {\bibfnamefont {E.}~\bibnamefont {Hall}}, \bibinfo {author} {\bibfnamefont {J.}~\bibnamefont {Read}},\ and\ \bibinfo {author} {\bibfnamefont {B.}~\bibnamefont {Sathyaprakash}},\ }\href {https://doi.org/10.48550/arXiv.2201.10668} {\bibinfo {title} {Science-driven tunable design of cosmic explorer detectors}} (\bibinfo {year} {2022})\BibitemShut {NoStop}%
\bibitem [{\citenamefont {{LIGO Scientific Collaboration}}(2024)}]{ligo_acronyms}%
  \BibitemOpen
  \bibfield  {author} {\bibinfo {author} {\bibnamefont {{LIGO Scientific Collaboration}}},\ }\href {https://dcc.ligo.org/LIGO-M080375/public} {\bibinfo {title} {{LIGO Laboratory Abbreviations and Acronyms}}} (\bibinfo {year} {2024})\BibitemShut {NoStop}%
\end{thebibliography}%

\clearpage
\appendix
\onecolumngrid

\section{Acronyms}\label{app:acronyms}
Acronyms for LIGO subsystems and analysis tools used in this study. Each acronym corresponds to a detector subsystem or data-analysis method referenced in the text and figures. A more expansive list of acronyms is provided at ~\cite{ligo_acronyms}.

    \begin{table}[h]
    \centering
    \begin{tabular}{l l}
    \hline\hline
    \textbf{Acronym} & \textbf{Definition} \\
    \hline
    ASC & Alignment Sensing and Control \\
    CAL & Calibration subsystem \\
    CNN & Convolutional Neural Network \\
    DIRECT & Deep Image Representation for Clustering of Transients \\
    GDS & Global Diagnostic System \\
    HPI & Hydraulic Pre-Isolator \\
    IMC & Input Mode Cleaner \\
    ISI & Internal Seismic Isolation \\
    LSC & Length Sensing and Control \\
    OMC & Output Mode Cleaner \\
    PEM & Physical Environment Monitoring \\
    PSL & Pre-Stabilized Laser \\
    SQZ & Squeezing subsystem \\
    SUS & Suspension subsystem \\
    TCS & Thermal Compensation System \\
    t-SNE & t-distributed Stochastic Neighbor Embedding \\
    \hline\hline
    \end{tabular}
    \caption{Acronyms for LIGO subsystems, analysis tools, and machine-learning methods used most often in this study.}
    \end{table}

\section{Channel Rankings}\label{app:ranks}
This appendix summarizes the highest-ranked auxiliary channels identified by OmegaNeuron for each case study presented in the main text. Channels are ranked by strain similarity between auxiliary channel spectrograms and the strain channel spectrogram in the GravitySpy feature space. For comparison, the corresponding Omega Scan correlation rankings are included. These tables support the subsystem-level coupling behavior across glitch classes.

\begin{table}
        \centering
        \resizebox{0.85\textwidth}{!}{%
        \begin{tabular}{l c c c}
        \hline\hline
        \textbf{Rank -- Channel} & \textbf{Subsystem} & \textbf{Strain Sim.} & \textbf{$\Omega$-Scan Rank} \\
        \hline
        1. CAL-DELTAL\_EXTERNAL\_DQ & CAL & 0.9992 & 5th \\
        2. OMC-DCPD\_SUM\_OUT\_DQ & OMC & 0.9989 & 4th \\
        3. CAL-DELTAL\_RESIDUAL\_DQ & CAL & 0.9987 & 2nd \\
        4. CAL-DARM\_ERR\_WHITEN\_OUT\_DQ & CAL & 0.9985 & 1st \\
        \hline\hline
        \end{tabular}
        }
        \caption{Top OmegaNeuron strain-similarity channels for GW150914 (no known glitch), compared against the corresponding Omega Scan correlation rankings. The agreement highlights OmegaNeuron’s ability to recover known unsafe channels (CAL, OMC) and demonstrates the metric’s reliability in identifying similarity even in the absence of a glitch.}
        \label{channels_150914}
\end{table}
    
\begin{table}
        \centering
        \resizebox{0.85\textwidth}{!}{%
        \begin{tabular}{l c c c}
        \hline\hline
        \textbf{Rank -- Channel} & \textbf{Subsystem} & \textbf{Strain Sim.} & \textbf{$\Omega$-Scan Rank} \\
        \hline
        1. LSC-POP\_A\_RF9\_Q\_ERR\_DQ & LSC & 0.9851 & 11th \\
        2. LSC-REFL\_A\_RF9\_Q\_ERR\_DQ & LSC & 0.9277 & 10th \\
        3. ASC-AS\_A\_RF36\_Q\_YAW\_OUT\_DQ & ASC & 0.7572 & 18th \\
        4. ASC-REFL\_B\_RF9\_I\_YAW\_OUT\_DQ & ASC & 0.6932 & 12th \\
        \hline\hline
        \end{tabular}
        }
        \caption{Top OmegaNeuron strain-similarity channels for a scattered-light glitch. The top-ranked channels originate primarily from the LSC and ASC subsystems, consistent with established optical scattering pathways observed in prior studies. Omega Scan rankings are included to show agreement between the two correlation approaches and to illustrate the improved prioritization provided by the similarity metric.}
        \label{channels_240620}
\end{table}
    
\begin{table}
        \centering
        \resizebox{0.85\textwidth}{!}{%
        \begin{tabular}{l c c c}
        \hline\hline
        \textbf{Rank -- Channel} & \textbf{Subsystem} & \textbf{Strain Sim.} & \textbf{$\Omega$-Scan Rank} \\
        \hline
        1. PSL-ISS\_SECONDLOOP\_SUM58\_REL\_OUT\_DQ & PSL & 0.9914 & 94th \\
        2. LSC-PRCL\_IN1\_DQ & LSC & 0.9905 & 63rd \\
        3. PEM-MX\_ACC\_BEAMTUBE\_1900X\_X\_DQ & PEM & 0.9892 & 59th \\
        4. LSC-PRCL\_OUT\_DQ & LSC & 0.9862 & 71st \\
        \hline\hline
        \end{tabular}
        }
        \caption{Top OmegaNeuron strain-similarity channels for the unclassified glitch at GPS 1403298875~s. The leading witness candidates arise from optical (PSL, LSC) and environmental (PEM) sensors. From this analysis we are able to narrow down potential witness channels. These rankings show consistent behavior across repeated occurrences of the glitch and demonstrate OmegaNeuron’s ability to identify plausible witness channels even for rare, single-event transients.}
        \label{channels_240624}
\end{table}

\end{document}